\documentclass[12pt]{article}

\usepackage{cite}
\usepackage{epsfig}
\usepackage{feynarts}

\makeatletter
\if@twoside \oddsidemargin -17pt \evensidemargin 00pt
\else \oddsidemargin 00pt \evensidemargin 00pt
\fi
\expandafter\ifx\csname mathrm\endcsname\relax\def\mathrm#1{{\rm #1}}\fi

\renewcommand{\theequation}{\thesection.\arabic{equation}}
\newcounter{saveeqn}

\@addtoreset{equation}{section}

\makeatother

\def\beq{\begin{equation}}
\def\eeq{\end{equation}}
\def\beqar{\begin{eqnarray}}
\def\eeqar{\end{eqnarray}}
\def\barr#1{\begin{array}{#1}}
\def\earr{\end{array}}
\def\bfi{\begin{figure}}
\def\efi{\end{figure}}
\def\btab{\begin{table}}
\def\etab{\end{table}}
\def\bce{\begin{center}}
\def\ece{\end{center}}
\def\nn{\nonumber}
\def\disp{\displaystyle}
\def\text{\textstyle}

\def\de{\delta}
\def\veps{\varepsilon}

\def\refeq#1{\mbox{(\ref{#1})}}
\def\reffi#1{\mbox{Fig.~\ref{#1}}}
\def\reffis#1{\mbox{Figs.~\ref{#1}}}
\def\reftas#1{\mbox{Tables~\ref{#1}}}
\def\refse#1{\mbox{Sect.~\ref{#1}}}

\def\citere#1{\mbox{Ref.~\cite{#1}}}
\def\citeres#1{\mbox{Refs.~\cite{#1}}}

\newcommand{\GeV}{\unskip\,\mathrm{GeV}}
\newcommand{\MeV}{\unskip\,\mathrm{MeV}}
\newcommand{\TeV}{\unskip\,\mathrm{TeV}}

\def\mathswitchr#1{\relax\ifmmode{\mathrm{#1}}\else$\mathrm{#1}$\fi}

\newcommand{\PW}{\mathswitchr W}
\newcommand{\PZ}{\mathswitchr Z}

\newcommand{\Pe}{\mathswitchr e}

\newcommand{\Pd}{\mathswitchr d}

\newcommand{\Pu}{\mathswitchr u}
\newcommand{\Ps}{\mathswitchr s}
\newcommand{\Pb}{\mathswitchr b}
\newcommand{\Pc}{\mathswitchr c}
\newcommand{\Pt}{\mathswitchr t}
\newcommand{\Pp}{\mathswitchr p}

\newcommand{\Pep}{\mathswitchr {e^+}}
\newcommand{\Pem}{\mathswitchr {e^-}}

\def\mathswitch#1{\relax\ifmmode#1\else$#1$\fi}

\newcommand{\MW}{\mathswitch {M_\PW}}

\newcommand{\MZ}{\mathswitch {M_\PZ}}

\newcommand{\Me}{\mathswitch {m_\Pe}}

\newcommand{\Md}{\mathswitch {m_\Pd}}
\newcommand{\Mu}{\mathswitch {m_\Pu}}
\newcommand{\Ms}{\mathswitch {m_\Ps}}
\newcommand{\Mc}{\mathswitch {m_\Pc}}
\newcommand{\Mb}{\mathswitch {m_\Pb}}
\newcommand{\Mt}{\mathswitch {m_\Pt}}

\newcommand{\scrs}{\scriptstyle}
\newcommand{\sw}{\mathswitch {s_{\scrs\PW}}}
\newcommand{\cw}{\mathswitch {c_{\scrs\PW}}}

\newcommand{\GF}{\mathswitch {G_\mu}}

\newcommand{\rd}{{\mathrm{d}}}

\newcommand{\soft}{{\mathrm{soft}}}

\newcommand{\M}{{\cal {M}}}

\newcommand{\ct}{{\mathrm{ct}}}
\newcommand{\virt}{{\mathrm{virt}}}
\newcommand{\coll}{{\mathrm{coll}}}
\newcommand{\sub}{{\mathrm{sub}}}

\def\Li{\mathop{\mathrm{Li}_2}\nolimits}

\def\Re{\mathop{\mathrm{Re}}\nolimits}

\def\draftdate{\relax}
\def\mda{\relax}
\def\mua{\relax}
\def\mla{\relax}
\def\draft{
\def\thtystars{******************************}
\def\sixtystars{\thtystars\thtystars}
\typeout{}
\typeout{\sixtystars**}
\typeout{* Draft mode!
         For final version remove \protect\draft\space in source file *}
\typeout{\sixtystars**}
\typeout{}
\def\draftdate{\today}
\def\mua{\marginpar[\boldmath\hfil$\uparrow$]%
                   {\boldmath$\uparrow$\hfil}%
                    \typeout{marginpar: $\uparrow$}\ignorespaces}
\def\mda{\marginpar[\boldmath\hfil$\downarrow$]%
                   {\boldmath$\downarrow$\hfil}%
                    \typeout{marginpar: $\downarrow$}\ignorespaces}
\def\mla{\marginpar[\boldmath\hfil$\rightarrow$]%
                   {\boldmath$\leftarrow $\hfil}%
                    \typeout{marginpar: $\leftrightarrow$}\ignorespaces}
\def\Mua{\marginpar[\boldmath\hfil$\Uparrow$]%
                   {\boldmath$\Uparrow$\hfil}%
                    \typeout{marginpar: $\Uparrow$}\ignorespaces}
\def\Mda{\marginpar[\boldmath\hfil$\Downarrow$]%
                   {\boldmath$\Downarrow$\hfil}%
                    \typeout{marginpar: $\Downarrow$}\ignorespaces}
\def\Mla{\marginpar[\boldmath\hfil$\Rightarrow$]%
                   {\boldmath$\Leftarrow $\hfil}%
                    \typeout{marginpar: $\Leftrightarrow$}\ignorespaces}
\overfullrule 5pt
\oddsidemargin -15mm
\marginparwidth 29mm
}

\makeatletter

\def\eqnarray{\stepcounter{equation}\let\@currentlabel=\theequation
\global\@eqnswtrue
\global\@eqcnt\z@\tabskip\@centering\let\\=\@eqncr
$$\halign to \displaywidth\bgroup\hskip\@centering
  $\displaystyle\tabskip\z@{##}$\@eqnsel&\global\@eqcnt\@ne
  \hskip 2\arraycolsep \hfil${##}$\hfil
  &\global\@eqcnt\tw@ \hskip 2\arraycolsep $\displaystyle\tabskip\z@{##}$\hfil
   \tabskip\@centering&\llap{##}\tabskip\z@\cr}
\def\appendix{\par
 \setcounter{section}{0} \setcounter{subsection}{0}
 \def\thesection{\Alph{section}}}

\def\dsl{\mathpalette\make@slash}
\def\make@slash#1#2{\setbox\z@\hbox{$#1#2$}%
  \hbox to 0pt{\hss$#1/$\hss\kern-\wd0}\box0}

\makeatother

\begin{document}

\thispagestyle{empty}
\def\thefootnote{\fnsymbol{footnote}}
\setcounter{footnote}{1}
\null
\draftdate\hfill Edinburgh 2003/05 \\
\strut\hfill MPI-PhT/2003-24 \\
\strut\hfill hep-ph/0306234

\vfill
\begin{center}
{\large \boldmath{\bf
Electroweak Radiative Corrections to \\[.5em]
Associated $WH$ and $ZH$ Production at Hadron Colliders}
\par} \vskip 2.5em
{\normalsize
{\sc M.L.\ Ciccolini$^1$, S.\ Dittmaier$^{2}$ and M.\ Kr\"amer$^1$%
}\\[3ex]
{\normalsize \it 
$^1$ School of Physics, The University~of~Edinburgh,
Edinburgh EH9 3JZ, Scotland \\[.2cm]
$^2$ Max-Planck-Institut f\"ur Physik (Werner-Heisenberg-Institut), 
D-80805 M\"unchen, Germany}}
\par \vskip 1em
\end{center}\par
\vskip .0cm \vfill \noindent{\bf Abstract:} \par \noindent
Higgs-boson production in association with $W$ or $Z$~bosons, $\,p\bar
p \to WH/ZH+X,$ is the most promising discovery channel for a light
Standard Model Higgs particle at the Fermilab Tevatron. We present
the calculation of the electroweak ${\cal O}(\alpha)$ corrections to
these processes. The corrections decrease the theoretical prediction
by up to 5--10\%, depending in detail on the Higgs-boson mass and the
input-parameter scheme. We update the cross-section prediction for
associated $WH$ and $ZH$ production at the Tevatron and at the LHC,
including the next-to-leading order electroweak and QCD corrections,
and study the theoretical uncertainties induced by factorization and 
renormalization scale dependences and by the
parton distribution functions.
\par
\vskip 1cm
\noindent
June 2003
\null
\setcounter{page}{0}
\clearpage
\def\thefootnote{\arabic{footnote}}
\setcounter{footnote}{0}

\section{Introduction}

The search for Higgs particles \cite{Higgs:1964ia} is one of the most
important endeavours for future high-energy collider
experiments. Direct searches at LEP have set a lower limit on the
Standard Model (SM) Higgs-boson mass of $M_H > 114.4\GeV$ at the 95\%
confidence level~(C.L.)~\cite{:2001xw}. SM analyses of electroweak
precision data, on the other hand, result in an upper limit of $M_H <
211\GeV$ at 95\%~C.L.~\cite{LEPEWWG2003}. The search for the Higgs
boson continues at the upgraded proton--antiproton collider
Tevatron~\cite{Carena:2000yx} with a centre-of-mass (CM) energy of
$1.96\TeV$, followed in the near future by the proton--proton collider
LHC~\cite{atlas_cms_tdrs} with $14\TeV$ CM energy. Various channels
can be exploited at hadron colliders to search for a Higgs boson. At
the Tevatron, Higgs-boson production in association with $W$ or
$Z$~bosons,
\begin{equation}
p\bar p \to WH+X \quad\mbox{and}\quad p\bar p \to ZH+X,
\label{eq:procs}
\end{equation}
is the most promising discovery channel for a SM Higgs
particle with a mass below about 135 GeV, where decays into
$b\bar{b}$ final states are dominant~\cite{Carena:2000yx}.  

At leading order, the production of a Higgs boson in association with
a vector boson, $p\bar p \to VH+X, (V=W,Z)$ proceeds through $q\bar{q}$
annihilation~\cite{Glashow:ab},
\begin{equation}
q\bar{q}' \to V^* \to V+H.
\label{eq:partprocs}
\end{equation}
The next-to-leading order (NLO) QCD corrections coincide with those
for the Drell-Yan process and increase the cross section by about
30\%~\cite{Han:1991ia}. Beyond NLO, the QCD corrections for $VH$
production differ from those for the Drell-Yan process by
contributions where the Higgs boson couples to a heavy fermion
loop. The impact of these additional terms is, however, expected to be
small in general~\cite{Dicus:1985wx}, and NNLO QCD corrections should
not increase the $VH$ cross section at the Tevatron significantly,
similar to the Drell-Yan cross section~\cite{Hamberg:1991np}. As
described in more detail in Section~\ref{se:numres}, the
renormalization and factorization scale dependence is reduced to about
$10\%$ at ${\cal O}(\alpha_{\rm s})$, while the uncertainty due to the
parton luminosity is less than about $5\%$. At this level of accuracy,
the electroweak ${\cal O}(\alpha)$ corrections become significant and
need to be included to further improve the theoretical prediction.
Moreover, the QCD uncertainties may be reduced by forming the ratios
of the associated Higgs-production cross section with the
corresponding Drell--Yan-like W- and Z-boson production channels,
i.e.\ by inspecting $\sigma_{p\bar{p}\to VH+X}/\sigma_{p\bar{p}\to
V+X}$.  In these ratios, higher-order electroweak effects should be
significant.  For the Drell--Yan-like W- and Z-boson production the
electroweak corrections have been calculated in
\citeres{Baur:1999kt,Dittmaier:2001ay} and
\cite{Baur:1997wa}, respectively.

In this paper we present the calculation of the electroweak ${\cal
O}(\alpha)$ corrections to the processes $p\bar p/pp \to W^+H+X$ and
$p\bar{p}/pp \to ZH+X$.%
\footnote{The electroweak ${\cal O}(\alpha)$
corrections to associated $ZH$ production at $e^+e^-$ colliders have
been presented in \citere{Fleischer:1982af}.}
We update the cross-section prediction for associated $WH$ and $ZH$
production at the Tevatron and at the LHC, including the NLO
electroweak and QCD corrections, and we quantify the residual
theoretical uncertainty due to scale variation and the parton
distribution functions.

The paper is organized as follows. In \refse{se:partoncs} we outline
the computation of the ${\cal O}(\alpha)$ electroweak corrections. The
calculation of the hadronic cross section and the treatment of the
initial-state mass singularities are described in \refse{se:ppcs}.  In
\refse{se:numres} we present numerical results for associated $WH$ and
$ZH$ production at the Tevatron and at the LHC. Our conclusions are
given in \refse{se:concl}.

\section{The parton cross section}
\label{se:partoncs}

\subsection{Conventions and lowest-order cross section}

\label{se:born}

We consider the parton process
\beq
q(p_q,\tau_q) + \bar{q}'(p_{\bar q'},\tau_{\bar{q}'}) \;\to\;
V(p_V,\lambda_V) + H(p_H) \;\; [+\gamma(k,\lambda)],
\eeq
where $V = W^+,Z$. The light up- and down-type quarks are denoted by
$q$ and $q'$, where $q=\Pu,\Pc$ and $q' =\Pd,\Ps$ for $W^+H$
production and $q=q'=\Pu,\Pd,\Ps,\Pc,\Pb$ for $ZH$ production. The
variables within parentheses refer to the momenta and helicities of
the respective particles.  The Mandelstam variables are defined by
\beq
\hat s = (p_q+p_{\bar q'})^2, \quad
\hat t = (p_q-p_V)^2, \quad
\hat u = (p_{\bar q'}-p_V)^2, \quad
s_{VH} = (p_H+p_V)^2.
\eeq
Obviously, we have $\hat s = s_{VH}$ for the non-radiative process
$q\bar q'\to V H$.  
We neglect the fermion masses $m_q$, $m_{\bar q'}$ whenever possible, i.e.\
we keep these masses only as regulators in the logarithmic mass
singularities originating from collinear photon emission or exchange.
As a consequence, the fermion helicities $\tau_q$ and $\tau_{\bar q'}$
are conserved in lowest order and in the virtual one-loop corrections,
i.e.\ the matrix elements vanish unless 
$\tau_q=-\tau_{\bar q'}\equiv\tau=\pm1/2$.
For brevity the value of $\tau$ is sometimes indicated by its sign.

In lowest order only the Feynman diagram shown in
\reffi{fig:borndiag} contributes to the scattering amplitude,
\begin{figure}
\centerline{\footnotesize  \input{qqvh.tree.tex}} \vspace*{-2em}
\caption{Lowest-order diagram for $q\bar q'\to V^*\to V H$ ($V=W,Z$).}
\label{fig:borndiag}
\end{figure}
and the corresponding Born matrix element is given by
\beq
\M_0^\tau = \frac{e^2 g^\tau_{qq'V} g_{VVH}}{\hat s-M_V^2} \;
\bar v(p_{\bar q'}) \, \dsl{\veps}^*_V(\lambda_V) \, \omega_\tau u(p_q),
\label{eq:m0}
\eeq
where $\veps^*_V(\lambda_V)$ is the polarization vector of the boson $V$,
$\bar v(p_{\bar q'})$ and $u(p_q)$ are the Dirac spinors of the quarks, and
$\omega_\pm=\frac{1}{2}(1\pm\gamma_5)$ denote the chirality projectors.
The coupling factors are given by
\beqar
g^\tau_{udW} &=& \parbox{14em}{$\disp \frac{V^*_{ud}}{\sqrt{2}\sw}\,\de_{\tau-},
\hfill \quad g_{WWH}$}
= \frac{\MW}{\sw},
\nn\\
g^\tau_{qqZ} &=& 
\parbox{14em}{$\disp -\frac{\sw}{\cw}Q_q+\frac{I_q^3}{\cw\sw}\,\de_{\tau-},
 \hfill\; g_{ZZH}$} = \frac{\MZ}{\cw\sw},
\eeqar
where $Q_q$ and $I_q^3=\pm1/2$ are the relative charge and the third 
component of the weak isospin of quark $q$, respectively. 
The weak mixing angle is fixed by the mass ratio $\MW/\MZ$,
according to the on-shell condition
$\sin^2 \theta_W\equiv s_W^2 =1-c_W^2 =1-\MW^2/\MZ^2$. 
Note that the CKM matrix element for the $ud$ transition, 
$V_{ud}$, appears only as global factor $|V_{ud}|^2$
in the cross section for $WH$ production, since corrections to
flavour mixing are negligible in the considered process.
This means that the CKM matrix is set to unity in the relative corrections
and, in particular, that the parameter $V_{ud}$ need not be renormalized.
The same procedure was already adopted for Drell--Yan-like W~production
\cite{Baur:1999kt,Dittmaier:2001ay}.

The differential lowest-order cross section is easily obtained by
squaring the lowest-order matrix element $\M^\tau_0$ of \refeq{eq:m0},
\beqar
\biggl(\frac{\rd\hat\sigma_0}{\rd\hat\Omega}\biggr)
& = & \frac{1}{12} \, \frac{1}{64\pi^2} \, 
\frac{\lambda^{1/2} (M_V^2, M_H^2, \hat{s})}{\hat s^2} \,
\sum_{\rm spins} |\M^\tau_0|^2 \nn \\
& = &
\frac{\alpha^2}{48\,M_V^2 \hat s^2} \, 
g_{VVH}^2 \left( (g_{qq'V}^+)^2 + (g_{qq'V}^-)^2 \right) \,
\lambda^{1/2} (M_V^2, M_H^2, \hat{s}) \nn \\
& & {} \times
\frac{(\hat t - M_V^2) (\hat u - M_V^2)+  M_V^2\hat{s}}{(\hat{s}-M_V^2)^2},
\label{eq:dcs}
\eeqar
where the explicit factor $1/12$ results from the average over the
quark spins and colours, and $\hat\Omega$ is the solid angle of the
vector boson $V$ in the parton CM frame.  The total parton cross
section is given by
\beqar
\hat{\sigma}_{0}(q\bar{q}' \to V H) &=& 
\frac{\alpha^2\pi}{72 M_V^2 \hat{s}^2} \,
g_{VVH}^2 \left( (g_{qq'V}^+)^2 + (g_{qq'V}^-)^2 \right)
\nn\\
&& {} \times
\lambda^{1/2} (M_V^2, M_H^2, \hat{s}) \,
\frac{\lambda(M_V^2, M_H^2, \hat{s})+12 M_V^2\hat{s}}{(\hat{s}-M_V^2)^2},
\eeqar
where $\lambda$ is the two-body phase space function
$\lambda(x,y,z)=x^2+y^2+z^2-2xy-2xz-2yz$. 
The electromagnetic
coupling $\alpha=e^2/(4\pi)$ can be set to different values according
to different input-parameter schemes.  It can be directly identified
with the fine-structure constant $\alpha(0)$ or the running
electromagnetic coupling $\alpha(k^2)$ at a high-energy scale $k$.
For instance, it is possible to make use of the value of
$\alpha(\MZ^2)$ that is obtained by analyzing
\cite{Jegerlehner:2001ca} the experimental ratio
$R=\sigma(\Pep\Pem\to\mbox{hadrons})/(\Pep\Pem\to\mu^+\mu^-)$.  These
choices are called {\it $\alpha(0)$-scheme} and {\it
$\alpha(\MZ^2)$-scheme}, respectively, in the following.  Another
value for $\alpha$ can be deduced from the Fermi constant $\GF$,
yielding $\alpha_{\GF}=\sqrt{2}\GF\MW^2\sw^2/\pi$; this choice is
referred to as {\it $\GF$-scheme}.  The differences between these
schemes will become apparent in the discussion of the corresponding
${\cal O}(\alpha)$ corrections.

\subsection{Virtual corrections}

\subsubsection{One-loop diagrams and calculational framework}

The virtual corrections can be classified into self-energy, vertex,
and box corrections. The generic contributions of the
different vertex functions are shown in \reffis{fi:wgendiag} and
\ref{fi:zgendiagrams}.
\begin{figure}
\centerline{\footnotesize  \input{udwh.gen.tex}} \vspace*{-1em}
\caption{Contributions of different vertex functions to 
$u\bar{d} \to WH$.}
\label{fi:wgendiag}
\vspace*{1em}
\centerline{\footnotesize  \input{qqzh.gen.tex}} \vspace*{-1em}
\caption{Contributions of different vertex functions to 
$q\bar q\to Z H$.}
\label{fi:zgendiagrams}
\end{figure}
Explicit results for the transverse parts of the 
$WW$, $ZZ$, and $\gamma Z$ self-energies
(in the `t~Hooft--Feynman gauge) can, e.g., be found in 
\citere{Denner:1993kt}.
The diagrams for the gauge-boson--fermion vertex corrections are
shown in \reffis{fi:udwvertdiag}, \ref{fi:xzvert}, and
\ref{fi:qqhvertdiag}.
\begin{figure}
\centerline{\footnotesize  \input{udw.vert.tex}} \vspace*{-1em}
\caption{Diagrams for the corrections to the
$u\bar{d}W$ vertex.}
\label{fi:udwvertdiag}
\vspace*{1em}
\centerline{\footnotesize  \input{uuz.vert.tex} } \vspace*{-1em}
\centerline{\footnotesize  \input{ddz.vert.tex} } \vspace*{-1em}
\caption{Diagrams for the corrections to the
$q\bar{q}Z$ vertices.}
\label{fi:xzvert}
\vspace*{1em}
\centerline{\footnotesize  \input{uuh.vert.tex} 
\input{ddh.vert.tex} } \vspace*{-1em}
\caption{Diagrams for the corrections to the
$q\bar{q}H$ vertices.}
\label{fi:qqhvertdiag}
\end{figure}
The diagrams for the corrections to the $WWH$, $ZZH$, and $\gamma ZH$
vertices can, e.g., be found in Figs.~8 and 9 of \citere{Denner:2003iy}.
The box diagrams are depicted in \reffis{fi:wboxdiagrams} and
\ref{fi:zboxdiagrams}, where $\varphi$ is the would-be Goldstone partner 
of the $W$ boson.
\begin{figure}
\centerline{\footnotesize  \input{udwh.box.tex}} \vspace*{-1em}
\caption{Diagrams for box corrections to $u\bar d\to WH$.}
\label{fi:wboxdiagrams}
\centerline{\footnotesize  \input{uuzh.box.tex}} \vspace*{-1em}
\centerline{\footnotesize  \input{ddzh.box.tex}} \vspace*{-1em}
\caption{Diagrams for box corrections to $q\bar q\to Z H$.}
\label{fi:zboxdiagrams}
\end{figure}

The actual calculation of the one-loop diagrams has been carried out
in the 't~Hooft--Feynman gauge using standard techniques. The Feynman
graphs have been generated with {\sl FeynArts}
\cite{Kublbeck:1990xc,Hahn:2000kx} 
and are evaluated in two completely independent ways, leading to two
independent computer codes.  The results of the two codes are in good
numerical agreement (i.e.\ within approximately 12
digits for non-exceptional phase-space points). In both calculations
ultraviolet divergences are regulated dimensionally and IR divergences
with an infinitesimal photon mass $m_\gamma$ and small quark
masses. The renormalization is carried out in the on-shell
renormalization scheme, as e.g.\ described in \citere{Denner:1993kt}.

In the first calculation, the Feynman graphs are generated with {\sl
FeynArts} version 1.0 \cite{Kublbeck:1990xc}. With the help of
{\sl Mathematica} routines the amplitudes are expressed in terms of
standard matrix elements, which contain the Dirac spinors and polarization
vectors, and coefficients of tensor integrals. 
The tensor coefficients are numerically reduced to scalar integrals
using the Passarino--Veltman algorithm \cite{Passarino:1979jh}. 
The scalar integrals are evaluated using the methods and results of
\citeres{'tHooft:1979xw,Denner:1993kt}.

The second calculation has been made using {\sl FeynArts} version~3
\cite{Hahn:2000kx} for the diagram generation and 
{\sl FeynCalc} version~4.1.0.3b
\cite{Mertig:1991an} for the algebraic manipulations of the amplitudes,
including the Passarino--Veltman reduction to scalar integrals.
The latter have been numerically evaluated using the {\sl Looptools}
package \cite{Hahn:2000jm} version~2.

\subsubsection{Renormalization and input-parameter schemes}
\label{se:ren_ips}

Denoting the one-loop matrix element $\M_1^\tau$,
in ${\cal O}(\alpha)$ the squared matrix element reads
\beq
|\M_0^\tau+\M_1^\tau|^2 = 
|\M_0^\tau|^2 + 2\Re\{(\M_0^\tau)^* \M_1^\tau\}+\dots = 
(1+2\Re\{\delta^\tau_{\virt}\}) |\M_0^\tau|^2 
+ \dots.
\label{eq:dm2virt}
\eeq
Substituting the r.h.s.\ of this equation for $|\M_0^\tau|^2$ in
\refeq{eq:dcs} includes the virtual corrections to the differential
parton cross section.  The full one-loop corrections are too lengthy
and untransparent to be reported completely. Instead we list the
relevant counterterms, all of which lead to contributions to
$\M_1^\tau$ that are proportional to the lowest-order matrix element
$\M_0^\tau$, $\M_{\ct}^\tau = \delta_{\ct}^\tau \M_0^\tau$.
Explicitly the counterterm factors $\delta_{\ct}^\tau$ for the
individual vertex functions read
\beqar
\delta_{\ct}^{WW} &=& -\de Z_W + \frac{\de\MW^2}{\hat s-\MW^2},
\nn\\
\delta_{\ct}^{udW,\tau} &=& \left(
\de Z_e - \frac{\de\sw}{\sw} + \frac{1}{2}\de Z_W
+\frac{1}{2}\de Z^-_u + \frac{1}{2}\de Z^-_d \right) \de_{\tau-},
\nn\\
\delta_{\ct}^{WWH} &=& \de Z_e - \frac{\de\sw}{\sw}
+\frac{\de\MW^2}{2\MW^2} + \de Z_W + \frac{1}{2}\de Z_H,
\nn\\
\delta_{\ct}^{ZZ} &=& -\de Z_{ZZ} + \frac{\de\MZ^2}{\hat s-\MZ^2},
\nn\\
\delta_{\ct}^{\gamma Z,\tau} &=& \frac{Q_q}{2g^\tau_{qqZ}}
\left[ \de Z_{ZA}\left(1-\frac{\MZ^2}{\hat s}\right)+\de Z_{AZ} \right],
\nn\\
\delta_{\ct}^{qqZ,\tau} &=& \de Z_e 
+ \frac{\de g^\tau_{qqZ}}{g^\tau_{qqZ}} + \frac{1}{2}\de Z_{ZZ}
+ \de Z^\tau_q - \frac{Q_q}{2g^\tau_{qqZ}}\de Z_{AZ},
\nn\\
\delta_{\ct}^{\gamma ZH,\tau} &=&  \frac{Q_q}{2g^\tau_{qqZ}} \,
\de Z_{ZA}\left(\frac{\MZ^2}{\hat s}-1\right),
\nn\\
\delta_{\ct}^{ZZH} &=& \de Z_e 
+ \frac{2\sw^2-\cw^2}{\cw^2}\frac{\de\sw}{\sw} + \frac{\de\MW^2}{2\MW^2}
+ \de Z_{ZZ} + \frac{1}{2}\de Z_H.
\eeqar
The index $\tau$ has been suppressed for those counterterms that do
not depend on the chirality. The explicit expressions for the
renormalization constants can, e.g., be found in \citere{Denner:1993kt}.
We merely focus on the charge renormalization constant $\de Z_e$ in 
the following. In the $\alpha(0)$-scheme
(i.e.\ the usual on-shell scheme) the electromagnetic coupling $e$
is deduced from the fine-structure constant $\alpha(0)$, as defined
in the Thomson limit. This fixes $\de Z_e$ to
\beq
\de Z_e\Big|_{\alpha(0)} = \frac{1}{2} \left.
\frac{\partial\Sigma^{AA}_{\mathrm{T}}(k^2)}{\partial k^2}\right|_{k^2=0}
-\frac{\sw}{\cw} \frac{\Sigma^{AZ}_{\mathrm{T}}(0)}{\MZ^2},
\eeq
with $\Sigma^{VV'}_{\mathrm{T}}(k^2)$ denoting the transverse part of
the $VV'$ gauge-boson self-energy with momentum transfer $k$.
In this scheme the charge renormalization
constant $\delta Z_e$ contains logarithms of the light-fermion masses,
inducing large corrections proportional to $\alpha\ln(m_f^2/\hat s)$,
which are related to the running of the electromagnetic coupling
$\alpha(k^2)$ from $k=0$ to a high-energy scale. In order to render
these quark-mass logarithms meaningful, it is necessary to adjust
these masses to the asymptotic tail of the hadronic contribution to
the vacuum polarization $\Pi^{AA}(k^2)\equiv\Sigma^{AA}_{\mathrm{T}}(k^2)/k^2$ 
of the photon.  Using
$\alpha(\MZ^2)$, as defined in \citere{Jegerlehner:2001ca}, as input
this adjustment is implicitly incorporated, and the charge renormalization
constant is modified to
\beq
\de Z_e\Big|_{\alpha(\MZ^2)} = \de Z_e\Big|_{\alpha(0)}
-\frac{1}{2}\Delta\alpha(\MZ^2),
\eeq
where 
\beq
\Delta\alpha(k^2) = \Pi^{AA}_{f\ne \Pt}(0)-\Re\{\Pi^{AA}_{f\ne \Pt}(k^2)\},
\eeq
\begin{sloppypar}
\noindent
with $\Pi^{AA}_{f\ne \Pt}$ denoting the photonic vacuum polarization
induced by all fermions other than the top quark (see also
\citere{Denner:1993kt}). In contrast to the $\alpha(0)$-scheme the 
coun\-ter\-term $\de Z_e|_{\alpha(\MZ^2)}$, and thus the whole relative
${\cal O}(\alpha)$ correction in the $\alpha(\MZ^2)$-scheme, 
does not involve
logarithms of light quark masses, since all corrections of the form
$\alpha^n\ln^n(m_f^2/\hat s)$ are absorbed in the lowest-order cross
section parametrized by
$\alpha(\MZ^2)=\alpha(0)/[1-\Delta\alpha(\MZ^2)]$.  In the
$\GF$-scheme, the transition from $\alpha(0)$ to $\GF$ is ruled by the
quantity $\Delta r$ \cite{Sirlin:1980nh,Denner:1993kt}, which is
deduced from muon decay,
\end{sloppypar}
\beq
\alpha_{\GF}=\frac{\sqrt{2}\GF\MW^2\sw^2}{\pi}
=\alpha(0)(1+\Delta r) \;+\; {\cal O}(\alpha^3).
\eeq
Therefore, the charge renormalization constant reads
\beq
\de Z_e\Big|_{\GF} = \de Z_e\Big|_{\alpha(0)}
- \frac{1}{2}\Delta r.
\eeq
Since $\Delta\alpha(\MZ^2)$ is explicitly contained in $\Delta r$, the
large fermion-mass logarithms are also resummed in the $\GF$-scheme. Moreover,
the lowest-order cross section in $\GF$-parametrization absorbs large 
universal corrections to the SU(2) gauge coupling $e/\sw$ 
induced by the $\rho$-parameter.

Finally, we consider the universal corrections related to the 
$\rho$-parameter, or more generally, the leading corrections 
induced by heavy top quarks in the loops. To this end, we have
extracted all terms in the corrections that are enhanced by 
a factor $\Mt^2/\MW^2$. These contributions
are conveniently expressed in terms of 
\beq
\Delta\rho_t = \frac{3\alpha}{16\pi\sw^2}\frac{\Mt^2}{\MW^2},
\eeq 
which is the leading contribution to the $\rho$ parameter.
For the various channels we obtain the following correction
factors to the cross sections in the $\GF$-scheme,
\beqar
\delta^{\mathrm{top}}_{q\bar q'\to WH} \Big|_{\GF}
&=& 2\delta_{WWH}^{\mathrm{top}}\Big|_{\GF},
\nn\\
\delta^{\mathrm{top}}_{q\bar q\to ZH} \Big|_{\GF}
&=& 2\delta_{ZZH}^{\mathrm{top}}\Big|_{\GF} + \Delta\rho_t\left[ 1-
   \frac{2Q_q\cw(g^+_{qqZ}+g^-_{qqZ})}
		{\sw[(g^+_{qqZ})^2+(g^-_{qqZ})^2]} 
        \right],
\eeqar
with
\beq
\delta_{WWH}^{\mathrm{top}}\Big|_{\GF} = 
\delta_{ZZH}^{\mathrm{top}}\Big|_{\GF} = 
-\frac{5}{6}\Delta\rho_t,
\eeq
in agreement with the results of \citere{Kniehl:1995at}.
As mentioned before, for $WH$ production the only effect of
$\Delta\rho_t$ is related to the $WWH$ vertex correction in the
$\GF$-scheme, while such corrections to the $q\bar q'W$ coupling are
entirely absorbed into the renormalized coupling $e/\sw$.

\subsection{Real-photon emission}

Real-photonic corrections are induced by the diagrams shown in
\reffis{fig:wbremdiags} and \ref{fig:zbremdiags}.
\bfi
\centerline{\footnotesize  \unitlength=1bp%

\begin{feynartspicture}(432,216)(4,2)

\FADiagram{}
\FAProp(0.,15.)(5.5,10.)(0.,){/Straight}{1}
\FALabel(2.18736,11.8331)[tr]{$u$}
\FAProp(0.,5.)(5.5,10.)(0.,){/Straight}{-1}
\FALabel(3.31264,6.83309)[tl]{$d$}
\FAProp(20.,17.)(15.5,13.5)(0.,){/Sine}{1}
\FALabel(17.2784,15.9935)[br]{$W$}
\FAProp(20.,10.)(15.5,13.5)(0.,){/ScalarDash}{0}
\FALabel(18.0681,12.2962)[bl]{$H$}
\FAProp(20.,3.)(12.,10.)(0.,){/Sine}{0}
\FALabel(15.4593,5.81351)[tr]{$\gamma$}
\FAProp(5.5,10.)(12.,10.)(0.,){/Sine}{-1}
\FALabel(8.75,8.93)[t]{$W$}
\FAProp(15.5,13.5)(12.,10.)(0.,){/ScalarDash}{1}
\FALabel(13.134,12.366)[br]{$\varphi$}
\FAVert(5.5,10.){0}
\FAVert(15.5,13.5){0}
\FAVert(12.,10.){0}

\FADiagram{}
\FAProp(0.,15.)(5.5,10.)(0.,){/Straight}{1}
\FALabel(2.18736,11.8331)[tr]{$u$}
\FAProp(0.,5.)(5.5,10.)(0.,){/Straight}{-1}
\FALabel(3.31264,6.83309)[tl]{$d$}
\FAProp(20.,17.)(15.5,13.5)(0.,){/Sine}{1}
\FALabel(17.2784,15.9935)[br]{$W$}
\FAProp(20.,10.)(15.5,13.5)(0.,){/ScalarDash}{0}
\FALabel(18.0681,12.2962)[bl]{$H$}
\FAProp(20.,3.)(12.,10.)(0.,){/Sine}{0}
\FALabel(15.4593,5.81351)[tr]{$\gamma$}
\FAProp(5.5,10.)(12.,10.)(0.,){/Sine}{-1}
\FALabel(8.75,8.93)[t]{$W$}
\FAProp(15.5,13.5)(12.,10.)(0.,){/Sine}{1}
\FALabel(13.134,12.366)[br]{$W$}
\FAVert(5.5,10.){0}
\FAVert(15.5,13.5){0}
\FAVert(12.,10.){0}

\FADiagram{}
\FAProp(0.,15.)(4.5,10.)(0.,){/Straight}{1}
\FALabel(1.57789,11.9431)[tr]{$u$}
\FAProp(0.,5.)(4.5,10.)(0.,){/Straight}{-1}
\FALabel(2.92211,6.9431)[tl]{$d$}
\FAProp(20.,17.)(13.,14.5)(0.,){/Sine}{1}
\FALabel(15.9787,16.7297)[b]{$W$}
\FAProp(20.,10.)(10.85,8.4)(0.,){/ScalarDash}{0}
\FALabel(18.5,10.52)[b]{$H$}
\FAProp(20.,3.)(13.,14.5)(0.,){/Sine}{0}
\FALabel(17.7665,4.80001)[tr]{$\gamma$}
\FAProp(4.5,10.)(10.85,8.4)(0.,){/Sine}{-1}
\FALabel(7.29629,8.17698)[t]{$W$}
\FAProp(13.,14.5)(10.85,8.4)(0.,){/ScalarDash}{1}
\FALabel(10.9431,11.9652)[r]{$\varphi$}
\FAVert(4.5,10.){0}
\FAVert(13.,14.5){0}
\FAVert(10.85,8.4){0}

\FADiagram{}
\FAProp(0.,15.)(4.5,10.)(0.,){/Straight}{1}
\FALabel(1.57789,11.9431)[tr]{$u$}
\FAProp(0.,5.)(4.5,10.)(0.,){/Straight}{-1}
\FALabel(2.92211,6.9431)[tl]{$d$}
\FAProp(20.,17.)(13.,14.5)(0.,){/Sine}{1}
\FALabel(15.9787,16.7297)[b]{$W$}
\FAProp(20.,10.)(10.85,8.4)(0.,){/ScalarDash}{0}
\FALabel(18.5,10.52)[b]{$H$}
\FAProp(20.,3.)(13.,14.5)(0.,){/Sine}{0}
\FALabel(17.7665,4.80001)[tr]{$\gamma$}
\FAProp(4.5,10.)(10.85,8.4)(0.,){/Sine}{-1}
\FALabel(7.29629,8.17698)[t]{$W$}
\FAProp(13.,14.5)(10.85,8.4)(0.,){/Sine}{1}
\FALabel(10.9431,11.9652)[r]{$W$}
\FAVert(4.5,10.){0}
\FAVert(13.,14.5){0}
\FAVert(10.85,8.4){0}

\FADiagram{}
\FAProp(0.,15.)(10.,5.5)(0.,){/Straight}{1}
\FALabel(3.19219,13.2012)[bl]{$u$}
\FAProp(0.,5.)(10.,13.)(0.,){/Straight}{-1}
\FALabel(3.17617,6.28478)[tl]{$d$}
\FAProp(20.,17.)(16.,13.5)(0.,){/Sine}{1}
\FALabel(17.4593,15.9365)[br]{$W$}
\FAProp(20.,10.)(16.,13.5)(0.,){/ScalarDash}{0}
\FALabel(17.6239,11.2517)[tr]{$H$}
\FAProp(20.,3.)(10.,5.5)(0.,){/Sine}{0}
\FALabel(14.6241,3.22628)[t]{$\gamma$}
\FAProp(10.,5.5)(10.,13.)(0.,){/Straight}{1}
\FALabel(11.07,9.25)[l]{$u$}
\FAProp(10.,13.)(16.,13.5)(0.,){/Sine}{-1}
\FALabel(12.8713,14.3146)[b]{$W$}
\FAVert(10.,5.5){0}
\FAVert(10.,13.){0}
\FAVert(16.,13.5){0}

\FADiagram{}
\FAProp(0.,15.)(10.,13.)(0.,){/Straight}{1}
\FALabel(5.30398,15.0399)[b]{$u$}
\FAProp(0.,5.)(10.,5.5)(0.,){/Straight}{-1}
\FALabel(5.0774,4.18193)[t]{$d$}
\FAProp(20.,17.)(15.5,13.5)(0.,){/Sine}{1}
\FALabel(17.2784,15.9935)[br]{$W$}
\FAProp(20.,10.)(15.5,13.5)(0.,){/ScalarDash}{0}
\FALabel(18.0681,12.2962)[bl]{$H$}
\FAProp(20.,3.)(10.,5.5)(0.,){/Sine}{0}
\FALabel(15.3759,5.27372)[b]{$\gamma$}
\FAProp(10.,13.)(10.,5.5)(0.,){/Straight}{1}
\FALabel(8.93,9.25)[r]{$d$}
\FAProp(10.,13.)(15.5,13.5)(0.,){/Sine}{-1}
\FALabel(12.8903,12.1864)[t]{$W$}
\FAVert(10.,13.){0}
\FAVert(10.,5.5){0}
\FAVert(15.5,13.5){0}

\end{feynartspicture}

} \vspace*{-1em}
\caption{Bremsstrahlung diagrams for 
$u\bar d\to WH+\gamma$.}
\label{fig:wbremdiags}
\vspace*{1em}
\centerline{\footnotesize  \unitlength=1bp%

\begin{feynartspicture}(216,108)(2,1)

\FADiagram{}
\FAProp(0.,15.)(10.,5.5)(0.,){/Straight}{1}
\FALabel(3.19219,13.2012)[bl]{$q$}
\FAProp(0.,5.)(10.,13.)(0.,){/Straight}{-1}
\FALabel(3.17617,6.28478)[tl]{$q$}
\FAProp(20.,17.)(16.,13.5)(0.,){/Sine}{0}
\FALabel(17.4593,15.9365)[br]{$Z$}
\FAProp(20.,10.)(16.,13.5)(0.,){/ScalarDash}{0}
\FALabel(17.6239,11.2517)[tr]{$H$}
\FAProp(20.,3.)(10.,5.5)(0.,){/Sine}{0}
\FALabel(14.6241,3.22628)[t]{$\gamma$}
\FAProp(10.,5.5)(10.,13.)(0.,){/Straight}{1}
\FALabel(11.07,9.25)[l]{$q$}
\FAProp(10.,13.)(16.,13.5)(0.,){/Sine}{0}
\FALabel(12.8713,14.3146)[b]{$Z$}
\FAVert(10.,5.5){0}
\FAVert(10.,13.){0}
\FAVert(16.,13.5){0}

\FADiagram{}
\FAProp(0.,15.)(10.,13.)(0.,){/Straight}{1}
\FALabel(5.30398,15.0399)[b]{$q$}
\FAProp(0.,5.)(10.,5.5)(0.,){/Straight}{-1}
\FALabel(5.0774,4.18193)[t]{$q$}
\FAProp(20.,17.)(15.5,13.5)(0.,){/Sine}{0}
\FALabel(17.2784,15.9935)[br]{$Z$}
\FAProp(20.,10.)(15.5,13.5)(0.,){/ScalarDash}{0}
\FALabel(18.0681,12.2962)[bl]{$H$}
\FAProp(20.,3.)(10.,5.5)(0.,){/Sine}{0}
\FALabel(15.3759,5.27372)[b]{$\gamma$}
\FAProp(10.,13.)(10.,5.5)(0.,){/Straight}{1}
\FALabel(8.93,9.25)[r]{$q$}
\FAProp(10.,13.)(15.5,13.5)(0.,){/Sine}{0}
\FALabel(12.8903,12.1864)[t]{$Z$}
\FAVert(10.,13.){0}
\FAVert(10.,5.5){0}
\FAVert(15.5,13.5){0}

\end{feynartspicture}

} \vspace*{-1em}
\caption{Bremsstrahlung diagrams for 
$q\bar q\to Z H+\gamma$.}
\label{fig:zbremdiags}
\efi
Helicity amplitudes for the processes $q\bar q'\to V H + \gamma$
($V=W,Z$) have been generated and evaluated using the program packages
{\sl MadGraph}~\cite{Stelzer:1994ta} and {\sl
HELAS}~\cite{Murayama:1992gi}. The result has been verified by an
independent calculation based on standard trace techniques.
The contribution $\hat\sigma_\gamma$ of the radiative process to the
parton cross section is given by
\beq
\hat\sigma_\gamma = \frac{1}{12}\frac{1}{2\hat s} \int \rd\Gamma_\gamma \,
\sum_{\rm spins} |\M_\gamma|^2,
\label{eq:hbcs}
\eeq
where the phase-space integral is defined by
\beq
\int \rd\Gamma_\gamma =
\int\frac{\rd^3 {\bf p}_H}{(2\pi)^3 2p_{H,0}}
\int\frac{\rd^3 {\bf p}_V}{(2\pi)^3 2p_{V,0}}
\int\frac{\rd^3 {\bf k}}{(2\pi)^3 2k_0} \,
(2\pi)^4 \delta(p_q+p_{\bar q'}-p_H-p_V-k).
\label{eq:dGg}
\eeq

\subsection{Treatment of soft and collinear singularities}
\label{se:virt+real}

The phase-space integral \refeq{eq:hbcs} diverges in the soft ($k_0\to
0$) and collinear ($p_q k, p_{\bar q'} k \to 0$) regions logarithmically if
the photon and fermion masses are set to zero. For the treatment of
the soft and collinear singularities we have applied the phase-space
slicing method. For associated $ZH$ production we have additionally
applied the dipole subtraction method. In the following
we briefly sketch these two approaches.

\subsubsection{Phase-space slicing}

Firstly, we made use of phase-space slicing, excluding the soft-photon and
collinear regions in the integral \refeq{eq:hbcs}.

In the soft-photon region $m_\gamma<k_0<\Delta E\ll\sqrt{\hat s}$
the bremsstrahlung cross section factorizes into the lowest-order 
cross section and a universal eikonal factor that depends on the
photon momentum $k$ (see e.g.\ \citere{Denner:1993kt}). Integration
over $k$ in the partonic CM frame yields a simple correction factor 
$\delta_{\soft}$ to the partonic Born cross section $\rd\hat\sigma_0$.
For $Z H$ production this factor is
\beq
\delta_{\soft} =
-\frac{\alpha}{\pi} Q_q^2 \left\{
   2\ln \biggl(\frac{2\Delta E}{m_\gamma}\biggr)
        \ln \biggl(\frac{m_q^2}{\hat s} \biggr)
   +2 \ln\biggl( \frac{2\Delta E}{m_\gamma}\biggr)
   + \frac{1}{2}\ln^2\biggl( \frac{m_q^2}{\hat s}\biggr)
   + \ln \biggl( \frac{m_q^2}{\hat s}\biggr) 
   + \frac{\pi^2}{3}
   \right\}
\label{eq:dsoftZH}.
\eeq
For $W H$ production the soft factor is
\beqar
\delta_{\soft} &=&
-\frac{\alpha}{2\pi}\left\{
Q_q^2 \left[
   2\ln \biggl(\frac{2\Delta E}{m_\gamma}\biggr)
        \ln \biggl(\frac{m_q^2}{\hat s} \biggr)
   +2 \ln\biggl( \frac{2\Delta E}{m_\gamma}\biggr)
   + \frac{1}{2}\ln^2\biggl( \frac{m_q^2}{\hat s}\biggr)
   + \ln \biggl( \frac{m_q^2}{\hat s}\biggr)
   + \frac{\pi^2}{3}
   \right] \right.
\nn\\ && \hspace*{2.5em} {}
+Q_{q'}^2 \left[
   2\ln \biggl(\frac{2\Delta E}{m_\gamma}\biggr)
        \ln \biggl(\frac{m_{q'}^2}{\hat s} \biggr)
   +2 \ln\biggl( \frac{2\Delta E}{m_\gamma}\biggr)
   + \frac{1}{2}\ln^2\biggl( \frac{m_{q'}^2}{\hat s}\biggr)
   + \ln \biggl( \frac{m_{q'}^2}{\hat s}\biggr)
   + \frac{\pi^2}{3}
   \right]
\nn\\ && \hspace*{2.5em} {}
+ \left[
   2\ln \biggl(\frac{2\Delta E}{m_\gamma}\biggr)
        \ln \biggl(\frac{\MW^2}{\hat s} \biggr)
   +2 \ln\biggl( \frac{2\Delta E}{m_\gamma}\biggr)
   + \frac{1}{2}\ln^2\biggl( \frac{\MW^2}{(p_W^0 + |{\bf{p}}_W| )^2}\biggr)
   \right.
\nn\\ && \hspace*{6.0em} {} \left.
   + \frac{p_W^0}{|{\bf{p}}_W|}\ln \biggl( \frac{\MW^2}{(p_W^0 + |{\bf{p}}_W| )^2}\biggr)
   \right]
\nn\\ && \hspace*{2.5em} {}
+ 2Q_q\left[
   2\ln \biggl( \frac{\hat s}{\MW^2 - \hat t}\biggr)
        \ln \biggl( \frac{2\Delta E}{m_\gamma}\biggr)
   + \Li\biggl( 1 + \frac{\sqrt{\hat s}}{\hat t -\MW^2} (p_W^0 - |{\bf{p}}_W|)\biggr)
   \right.
\nn\\ && \hspace*{6.0em} {} \left.
   + \Li\biggl( 1 + \frac{\sqrt{\hat s}}{\hat t -\MW^2} (p_W^0 + |{\bf{p}}_W|)\biggr)
   \right]
\nn\\ && \hspace*{2.5em} {}
- 2Q_{q'}\left[
   2\ln \biggl( \frac{\hat s}{\MW^2 - \hat u}\biggr)
        \ln \biggl( \frac{2\Delta E}{m_\gamma}\biggr)
   + \Li\biggl( 1 + \frac{\sqrt{\hat s}}{\hat u -\MW^2} (p_W^0 - |\bf{p}_W|)\biggr)
   \right.
\nn\\ && \hspace*{6.0em} {} \left.\left.
   + \Li\biggl( 1 + \frac{\sqrt{\hat s}}{\hat u -\MW^2} (p_W^0 + |\bf{p}_W|)\biggr)
   \right]
\right\}, 
\label{eq:dsoft}
\eeqar
where $Q_q-Q_{q'}=+1$.
The difference between $ZH$ and $WH$ production is due to the soft photons
emitted by the $W$ boson.

The factor $\delta_{\soft}$ can be added directly to the virtual
correction factor $2\Re\{\delta^\tau_{\virt}\}$ defined in
\refeq{eq:dm2virt}. We have checked that the photon mass 
$m_\gamma$ cancels in the sum $2\Re\{\delta^\tau_{\virt}\}+\delta_{\soft}$.

The remaining phase-space integration in \refeq{eq:hbcs} with
$k_0>\Delta E$ still contains the collinear singularities in the
regions in which $(p_q k)$ or $(p_{\bar q'} k)$ is small. 
Defining $\theta_{f\gamma}=\angle({\bf p}_f,{\bf k})$ as the angle
of the photon emission off $f=q,\bar q'$, the collinear
regions are excluded by the angular cuts
$\theta_{f\gamma}<\Delta\theta\ll 1$ in the integral \refeq{eq:hbcs}.

In the collinear cones the photon emission angles $\theta_{f\gamma}$
can be integrated out. The resulting contribution to the 
bremsstrahlung cross section has the form of a convolution of the
lowest-order cross section,
\beqar
\hat\sigma_{\coll} &=& \hat\sigma_{\coll,q} + \hat\sigma_{\coll,\bar q'}, 
\nn\\[.5em]
\hat\sigma_{\coll,f}(p_f) &=&
\frac{Q_f^2 \alpha}{2\pi} \int_0^{1-2\Delta E/\sqrt{\hat s}} \rd z\,
\left[ \ln \left( \frac{\Delta\theta^2 \hat s}{4m_f^2} \right) - 
\frac{2z}{1+z^2}\right]
P_{ff}(z) \hat\sigma_0(zp_f),
\qquad f=q,\bar q'
\nn\\
\label{eq:qcoll}
\eeqar
with the splitting function
\beq
P_{ff}(z) = \frac{1+z^2}{1-z}.
\eeq
Note that the quark momentum $p_f$ is reduced by the factor $z$ so
that the partonic CM frame for the hard scattering receives a boost.

\subsubsection{Subtraction method}
\label{se:sub}

\begin{sloppypar}
Alternatively, for $ZH$ production, we applied the subtraction method
presented in \citere{Dittmaier:2000mb}, where the so-called ``dipole
formalism'', originally introduced by Catani and Seymour
\cite{Catani:1996jh} within massless QCD, was applied to photon
radiation and generalized to massive fermions.  The general idea of a
subtraction method is to subtract and to add a simple auxiliary
function from the singular integrand.  This auxiliary function has to
be chosen such that it cancels all singularities of the original
integrand so that the phase-space integration of the difference can be
performed numerically.  Moreover, the auxiliary function has to be
simple enough so that it can be integrated over the singular regions
analytically, when the subtracted contribution is added again.
\end{sloppypar}

The dipole subtraction function consists of contributions
labelled by all ordered pairs of charged external particles, one of
which is called {\it emitter}, the other one {\it spectator}.
For $q \bar q \to Z H$ we, thus, have 2 different 
emitter/spectator cases $ff'$: $q\bar q$, $\bar qq$.
The subtraction function that is subtracted from 
$\sum_{\rm spins}|\M_\gamma|^2$ is given by
\beqar
|\M_{\sub}|^2 &=& 
Q_q^2 e^2 \Bigl[ g_{q\bar q}(p_q,p_{\bar q},k) 
\sum_{\rm spins}|\M_0^{\tau}(x_{q\bar q}p_q,p_{\bar q},k_{Z,q \bar q})|^2
\nn\\
&& \hspace*{4.5em} {}
+g_{\bar q q}(p_{\bar q},p_q,k) \sum_{\rm spins}|\M_0^\tau(p_q,x_{q\bar q}p_{\bar q},k_{Z,\bar q q})|^2 \Bigr]
\label{eq:msub}
\eeqar
with the functions
\beqar
g_{q \bar q}(p_q,p_{\bar q},k) &=& \frac{1}{(p_q k)x_{q\bar q}}
\left[ \frac{2}{1-x_{q\bar q}}-1-x_{q\bar q} \right],
\nn\\
g_{\bar q q}(p_{\bar q},p_q,k) &=& \frac{1}{(p_{\bar q} k)x_{q\bar q}}
\left[ \frac{2}{1-x_{q\bar q}}-1-x_{q\bar q} \right],
\eeqar
and the auxiliary variable
\beq
x_{q \bar q} = \frac{p_q p_{\bar q}-p_q k-p_{\bar q} k}{p_q p_{\bar q}}, \qquad
\eeq
For the evaluation of $|\M_{\sub}|^2$ in \refeq{eq:msub} the $Z$-boson
momenta $k_{Z,ff'}$ still have to be specified.
They are given by
\beq
k_{Z,q \bar q}^\mu = {\Lambda(p_q,p_{\bar q})^\mu}_\nu k_Z^\nu, \qquad
k_{Z,\bar q q}^\mu = {\Lambda(p_{\bar q},p_q)^\mu}_\nu k_Z^\nu,
\eeq
with the Lorentz transformation matrix
\beq
{\Lambda(p_1,p_2)^\mu}_\nu = 
{g^\mu}_\nu - \frac{(P+\tilde P)^\mu(P+\tilde P)_\nu}{P^2+P\tilde P}
+\frac{2\tilde P^\mu P_\nu}{P^2}, \qquad
P^\mu=k_Z^\mu+k_H^\mu, \quad \tilde P^\mu=x_{q \bar q}p_1^\mu+p_2^\mu.
\eeq
The modified $Z$-boson momenta $k_{Z,ff'}$ still
obey the on-shell condition $k_{Z,ff'}^2=\MZ^2$,
and the same is true for the corresponding
Higgs-boson momenta that result from momentum conservation.  It is
straightforward to check that all collinear and soft singularities
cancel in $\sum_{\rm spins}|\M_\gamma|^2-|\M_{\sub}|^2$ so that
this difference can be integrated numerically over the entire phase
space \refeq{eq:dGg}.

The contribution of $|\M_{\sub}|^2$, which has been subtracted by
hand, has to be added again. This is done after the singular degrees
of freedom in the phase space \refeq{eq:dGg} are integrated out
analytically, keeping an infinitesimal photon mass $m_\gamma$ and
small fermion masses $m_f$ as regulators \cite{Dittmaier:2000mb}.  The
resulting contribution is split into two parts: one that factorizes
from the lowest-order cross section $\hat\sigma_0$ and another part that
has the form of a convolution integral over $\hat\sigma_0$ with reduced CM
energy.  The first part is given by
\beqar
\hat\sigma_{\sub,1} &=& 
Q_q^2 \frac{\alpha}{2\pi} 
\left[ 2{\cal L}( \hat s,m_q^2) 
	+3-\frac{2\pi^2}{3} \right]
\hat\sigma_0
\eeqar
with the auxiliary function
\beq
{\cal L}(r,m_q^2) =
\ln\biggl(\frac{m_q^2 }{r}\biggr)
\ln\biggl(\frac{m_\gamma^2}{r}\biggr)
+ \ln\biggl(\frac{m_\gamma^2}{r}\biggr)
- \frac12\ln^2\biggl(\frac{m_q^2}{r}\biggr)
+ \frac12\ln\biggl(\frac{m_q^2}{r}\biggr).
\label{eq:L}
\eeq
The IR and fermion-mass singularities contained in
$\rd\hat\sigma_{\sub,1}$ exactly cancel those of the virtual
corrections.  The second integrated subtraction contribution is given
by
\beqar
\hat\sigma_{\sub,2}(p_q,p_{\bar q}) &=&
Q_q^2 \frac{\alpha}{2\pi} \int_0^1 \rd x\, \left\{
 \left[ {\cal G}_{q \bar q}(\hat s,x) \right]_+ \hat\sigma_0(xp_q,p_{\bar q})
+\left[ {\cal G}_{\bar q q}(\hat s,x) \right]_+ \hat\sigma_0(p_q,xp_{\bar q}) \right\}
\Big|_{\hat s=(p_q+p_{\bar q})^2}
\nn\\ && {} 
\nn\\ && {} 
+ Q_q^2 \frac{\alpha}{2\pi} \int_0^1 \rd x\, (1-x)
\left\{ \hat\sigma_0(xp_q,p_{\bar q})\Big|_{\tau_q\to -\tau_q}
        + \hat\sigma_0(p_q,xp_{\bar q})\Big|_{\tau_{\bar q}\to -\tau_{\bar q}} \right\},
\nn\\
\label{eq:subconv}
\eeqar
where the usual $[\dots]_+$ prescription,
\beq
\int_0^1\rd x\, \Big[f(x)\Big]_+ g(x) =
\int_0^1\rd x\, f(x) \left[g(x)-g(1)\right],
\eeq
is applied to the integration kernels
\beq
{\cal G}_{q\bar q}(r,x) = 
{\cal G}_{\bar qq}(r,x) = 
P_{ff}(x) \left[ \ln\left(\frac{r}{m_q^2}\right)-1 \right].
\eeq
In \refeq{eq:subconv} we have indicated explicitly how the Mandelstam
variable $r$ has to be chosen in terms of the momenta in the
evaluation of the part containing $[{\cal G}_{ff'}(r,x)]_+$.  Note,
however, that in \refeq{eq:subconv} the variable $\hat s$ that is
implicitly used in the calculation of $\hat\sigma_0(\dots)$ is reduced
to $2xp_q p_{\bar q}=x\hat s$.

In summary, within the subtraction approach the real correction reads
\beq
\hat\sigma_\gamma = \frac{1}{12}\frac{1}{2\hat s} \int \rd\Gamma_\gamma \,
\left[ \sum_{\rm spins} |\M_\gamma|^2-|\M_{\sub}|^2 \right]
+ \hat\sigma_{\sub,1} + \hat\sigma_{\sub,2}.
\label{eq:sigmasub}
\eeq
It should be realized that in $\hat\sigma_{\sub,1}$ and $\hat\sigma_{\sub,2}$
the full photonic phase space is integrated over. This does, however,
not restrict the subtraction approach to observables that are fully
inclusive with respect to emitted photons, but rather to observables that 
are inclusive with respect to photons that are soft or collinear to any
charged external fermion (see discussions in Sect.~6.2 of 
\citere{Dittmaier:2000mb} and Sect.~7 of \citere{Catani:1996jh}).

\section{The hadron cross section} 
\label{se:ppcs}

The proton--(anti-)proton cross section $\sigma$ is obtained from the
parton cross sections $\hat\sigma^{(q_1 q_2)}$ by convolution with the
corresponding parton distribution functions $q_{1,2}(x)$,
\beq
\rd\sigma(s) = \sum_{q_1 q_2} \int_0^1 \rd x_1 \, \int_0^1 \rd x_2 \,
q_1(x_1) q_2(x_2) \,\rd\hat\sigma^{(q_1 q_2)}(p_{q_1}, p_{q_2}),
\label{eq:sigpp}
\eeq
where $x_{1,2}$ are the respective momentum fractions carried by the partons $q_{1,2}$.
In the sum $\sum_{q_1 q_2}$ the quark pairs $q_1 q_2$ run over all
possible combinations $q \bar q'$ and $\bar q' q$ where $q=\Pu,\Pc$
and $q' =\Pd,\Ps$ for $WH$ production and $q=q'=\Pu,\Pd,\Ps,\Pc,\Pb$
for $ZH$ production.  The squared CM energy $s$ of the $\Pp\Pp$
($\Pp\bar\Pp$) system is related to the squared parton CM energy $\hat
s$ by $\hat s=x_1 x_2 s$.

The ${\cal O}(\alpha)$-corrected parton cross section
$\hat\sigma^{(q_1 q_2)}$ contains mass singularities of the form
$\alpha\ln(m_q)$, which are due to collinear photon radiation off the
initial-state quarks.  In complete analogy to the
$\overline{\mbox{MS}}$ factorization scheme for next-to-leading order
QCD corrections, we absorb these collinear singularities into the
quark distributions.  This is achieved by replacing $q(x)$ in
\refeq{eq:sigpp} according to 
\beqar
q(x) &\to& q(x,M^2) 
\nn\\
&& {} 
-\int_x^1 \frac{\rd z}{z} \, q\biggl(\frac{x}{z},M^2\biggr) \,
\frac{\alpha}{2\pi} \, Q_q^2 \, \biggl\{
\ln\biggl(\frac{M^2}{m_q^2}\biggr) \Bigl[ P_{ff}(z) \Bigr]_+
-\Bigl[ P_{ff}(z) \Bigl(2\ln(1-z)+1\Bigr) \Bigr]_+
\biggr\},
\nn\\
\label{eq:factorization}
\eeqar
where $M$ is the factorization scale (see \citere{Dittmaier:2001ay}). 
This replacement defines the
same finite parts in the ${\cal O}(\alpha)$ correction as the usual 
$\overline{\mbox{MS}}$ factorization in $D$-dimensional regularization
for exactly massless partons, where the $\ln(m_q)$ terms appear as
$1/(D-4)$ poles.
In \refeq{eq:factorization}
we have regularized the
soft-photon pole by using the $[\dots]_+$
prescription. This procedure is fully equivalent to the application of
a soft-photon cutoff $\Delta E$ (see~\citere{Baur:1999kt}) where 
\begin{eqnarray}\label{eq:factorization-cut}
q(x) &\to&
 q(x,M^2) \; \Biggl[ 1 - \frac{\alpha}{\pi} \; Q_q^2 
\Biggl\{1-\ln(2\Delta E/\sqrt{\hat{s}}) -\ln^2(2\Delta E/\sqrt{\hat{s}})
\nonumber\\ 
&& \hspace*{18mm} + \left(\ln(2\Delta E/\sqrt{\hat{s}})
+\frac{3}{4}\right)\,\ln\left(\frac{M^2}{m_q^2}\right)\Biggr\}\Biggr]
\nonumber\\[2.mm]
&-& \int_x^{1-2\Delta E/\sqrt{\hat{s}}} \frac{d z}{z}\; 
q\left(\frac{x}{z},M^2\right)
\; \frac{\alpha}{2 \pi} \, Q_q^2 \,
P_{ff}(z)\left\{
\ln\left(\frac{M^2}{m_q^2}\frac{1}{(1-z)^2}\right)
-1\right\}.
\end{eqnarray}

The absorption of the collinear singularities of ${\cal O}(\alpha)$
into quark distributions, as a matter of fact, requires also the
inclusion of the corresponding ${\cal O}(\alpha)$ corrections into the
DGLAP evolution of these distributions and into their fit to
experimental data. At present, this full incorporation of ${\cal
O}(\alpha)$ effects in the determination of the quark distributions
has not yet been performed. However, an approximate inclusion of the
${\cal O}(\alpha)$ corrections to the DGLAP evolution shows
\cite{Kripfganz:1988bd} that the impact of these corrections on the
quark distributions in the $\overline{\mbox{MS}}$ factorization scheme
is well below 1\%, at least in the $x$ range that is relevant for
associated $VH$ production at the Tevatron and the LHC.  Therefore,
the neglect of these corrections to the parton distributions is
justified for the following numerical study.

\section{Numerical results}
\label{se:numres}

\subsection{Input parameters}

For the numerical evaluation we used the following set of input
parameters~\cite{Hagiwara:pw},
\beq
\begin{array}[b]{rclrclrcl}
\GF & = & 1.16639 \times 10^{-5} \GeV^{-2}, \quad&
\alpha(0) &=& 1/137.03599976,\quad & \alpha(M_Z) & = & 1/128.930, \\
\MW & = & 80.423\GeV, &
\MZ & = & 91.1876\GeV, & & &\\
\Me & = & 0.510998902\MeV, &
m_\mu &=& 105.658357\MeV, &
m_\tau &=& 1.77699\GeV, \\
\Mu & = & 66\MeV, &
\Mc & = & 1.2\GeV, &
\Mt & = & 174.3\;\GeV, \\
\Md & = & 66\MeV, &
\Ms & = & 150\MeV, &
\Mb & = & 4.3\GeV\\
|V_{\Pu\Pd}| & = & 0.975, &
|V_{\Pu\Ps}| & = & 0.222, \\
|V_{\Pc\Pd}| & = & 0.222, &
|V_{\Pc\Ps}| & = & 0.975.
\end{array}
\label{eq:SMpar}
\eeq
The masses of the light quarks are adjusted such as to reproduce the
hadronic contribution to the photonic vacuum polarization
of~\citere{Jegerlehner:2001ca}.  They are relevant only for the
evaluation of the charge renormalization constant $\delta Z_e$ in the
$\alpha(0)$-scheme.  For the calculation of the $\Pp\Pp$ and
$\Pp\bar\Pp$ cross sections we have adopted the CTEQ6L1 and CTEQ6M
\cite{Pumplin:2002vw} parton distribution functions at LO and ${\cal
O}(\alpha_{\rm s})$, 
corresponding to $\Lambda_5^{\mathrm{LO}}=165\MeV$ and
$\Lambda_5^{\overline{\mathrm{MS}}}=226\MeV$ at the one- and two-loop
level of the strong coupling $\alpha_{\mathrm{s}}(\mu)$, respectively.
The top quark is 
decoupled from the running of $\alpha_{\mathrm{s}}(\mu)$. If not
stated otherwise the factorization scale $M$ is set to the invariant
mass of the Higgs--vector-boson pair, $M = \sqrt{s_{VH}}$.  For the
treatment of the soft and collinear singularities we have applied the
phase-space slicing method as described in
\refse{se:virt+real}. We have verified that the results are independent 
of the slicing parameters $2\Delta E/\sqrt{\hat s}$ and $\Delta
\theta$ when these parameters are varied within the range $10^{-2}-10^{-4}$.
In the case of associated $ZH$ production we have, in addition,
applied the dipole subtraction method. The results agree with those
obtained using phase-space slicing. We observe that the integration
error of the subtraction method is smaller than that of the slicing
method by at least a factor of two.

\subsection{Electroweak corrections}\label{sec:ewk}

\begin{figure}[p]
\begin{center}

\vspace*{-5mm}

\epsfig{file=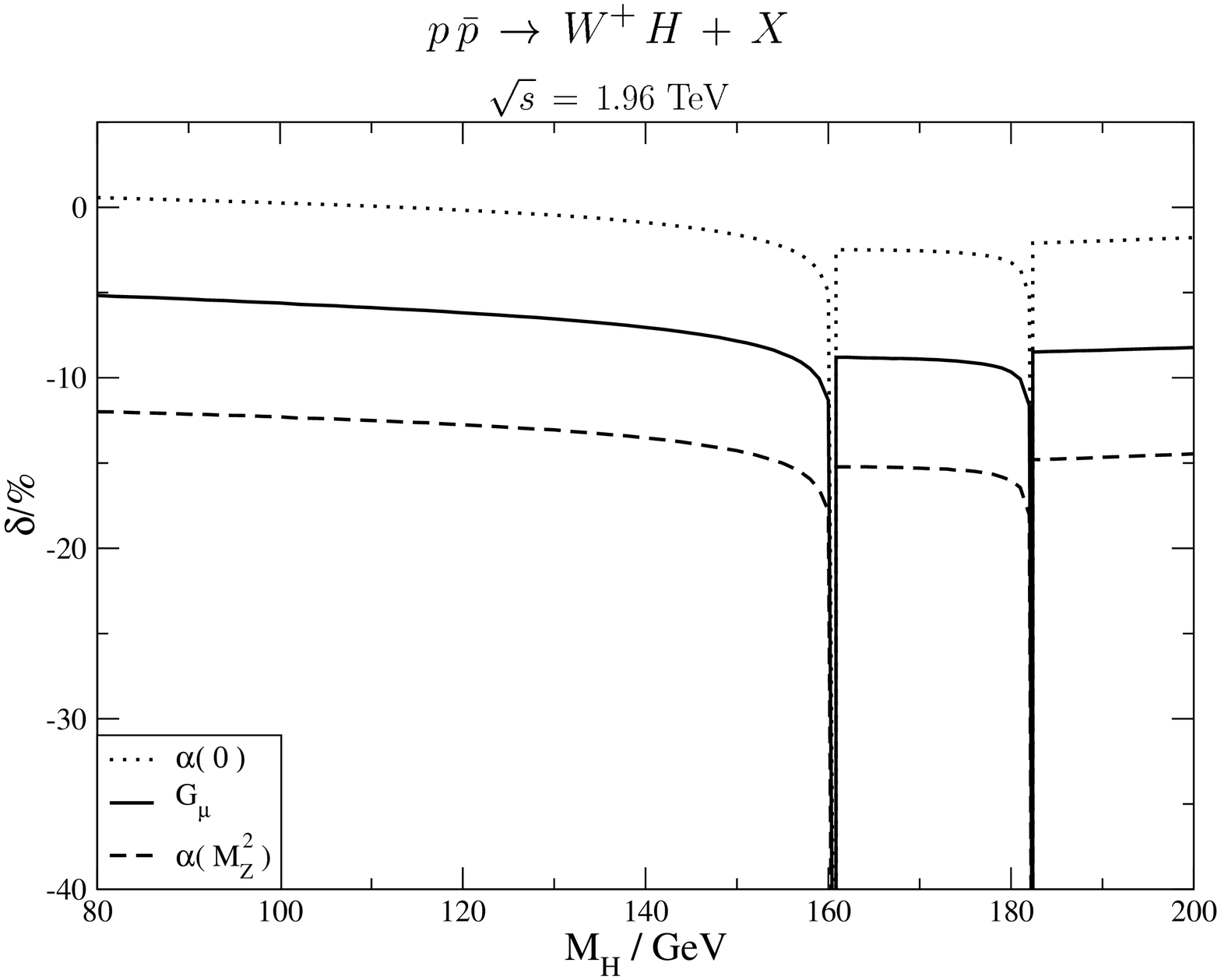,%
        bbllx=35pt,bblly=50pt,bburx=719pt,bbury=582pt,scale=0.5}
\vspace*{5mm}
\caption{\label{fig:tevwh}Relative electroweak correction $\delta$ as
 a function of the Higgs-boson mass for the total cross section
 $p\bar{p}\to W^+ H + X$ ($\sqrt{s}=1.96\;\mathrm{TeV}$). Results 
 are presented for the $\alpha(0)$-, $\alpha(\MZ^2)$-, and 
 $\GF$-schemes.}

\vspace*{5mm}

\epsfig{file=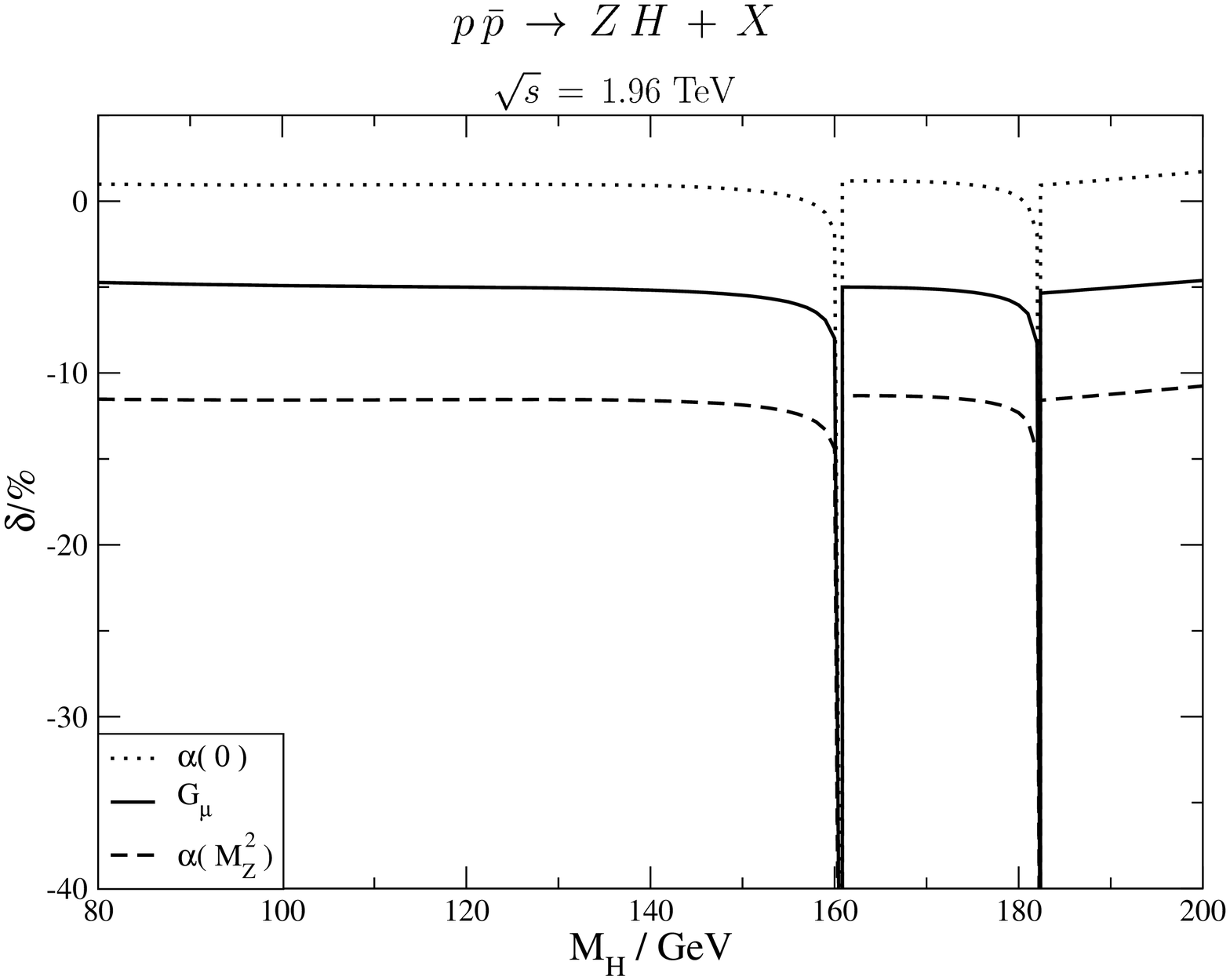,%
        bbllx=35pt,bblly=50pt,bburx=719pt,bbury=582pt,scale=0.5}
\vspace*{5mm}
\caption{\label{fig:tevzh}Relative electroweak correction $\delta$ as
 a function of the Higgs-boson mass for the total cross section
 $p\bar{p}\to ZH + X$ ($\sqrt{s}=1.96\;\mathrm{TeV}$).}
\end{center}
\end{figure}

In this subsection we present the impact of the electroweak ${\cal
O}(\alpha)$ corrections on the cross section predictions for the
processes $p\bar p/pp \to W^+H+X$ and $p\bar{p}/pp \to ZH+X$ at the
Tevatron and the LHC. Figures~\ref{fig:tevwh} and \ref{fig:tevzh} show
the relative size of the ${\cal O}(\alpha)$ corrections as a function
of the Higgs-boson mass for $p\bar p \to W^+ H + X$ and $p\bar p \to
ZH + X$ at the Tevatron.  Results are presented for the three
different input-parameter schemes. The corrections in the $\GF$- and
$\alpha(\MZ^2)$-schemes are significant and reduce the cross section
by 5--9\% and by 10--15\%, respectively. The corrections in the
$\alpha(0)$-scheme differ from those in the $\GF$-scheme by $2\Delta
r\approx 6\%$ and from those in the $\alpha(\MZ^2)$-scheme by
$2\Delta\alpha(\MZ^2)\approx 12\%$. The fact that the relative
corrections in the $\alpha(0)$-scheme are rather small results from
accidental cancellations between the running of the electromagnetic
coupling, which leads to a contribution of about
$2\Delta\alpha(\MZ^2)\approx +12\%$, and other (negative) corrections
of non-universal origin.  Thus, corrections beyond ${\cal O}(\alpha)$
in the $\alpha(0)$-scheme cannot be expected to be suppressed as well.
In all schemes, the size of the corrections does not depend strongly
on the Higgs-boson mass. The unphysical singularities at the
thresholds $M_H = 2M_W$ and $2M_Z$ can be removed by taking into
account the finite widths of the unstable particles, see e.g.\
\citeres{Bhattacharya:1991gr}. Representative results for the leading-order 
cross section and the electroweak ${\cal O}(\alpha)$ corrections are
collected in \reftas{tab:tevwh} and \ref{tab:tevzh}.

\begin{figure}[p]
\begin{center}

\vspace*{-5mm}

\epsfig{file=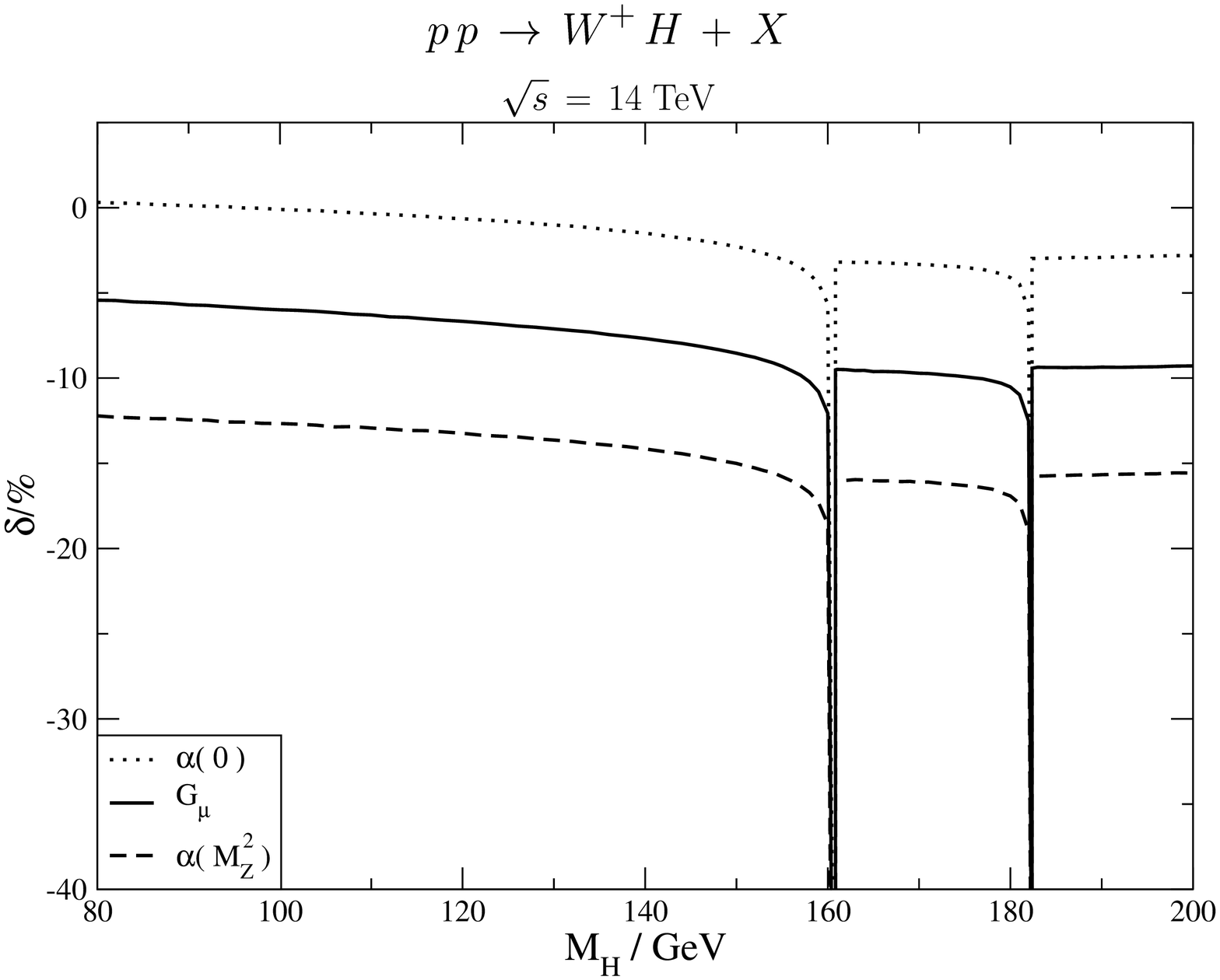,%
        bbllx=35pt,bblly=50pt,bburx=719pt,bbury=582pt,scale=0.5}
\vspace*{5mm}
\caption{\label{fig:lhcwh} Relative electroweak correction $\delta$ as
 a function of the Higgs-boson mass for the total cross section $pp\to
 W^+ H + X$ ($\sqrt{s}=14\;\mathrm{TeV}$).}

\vspace*{5mm}

\epsfig{file=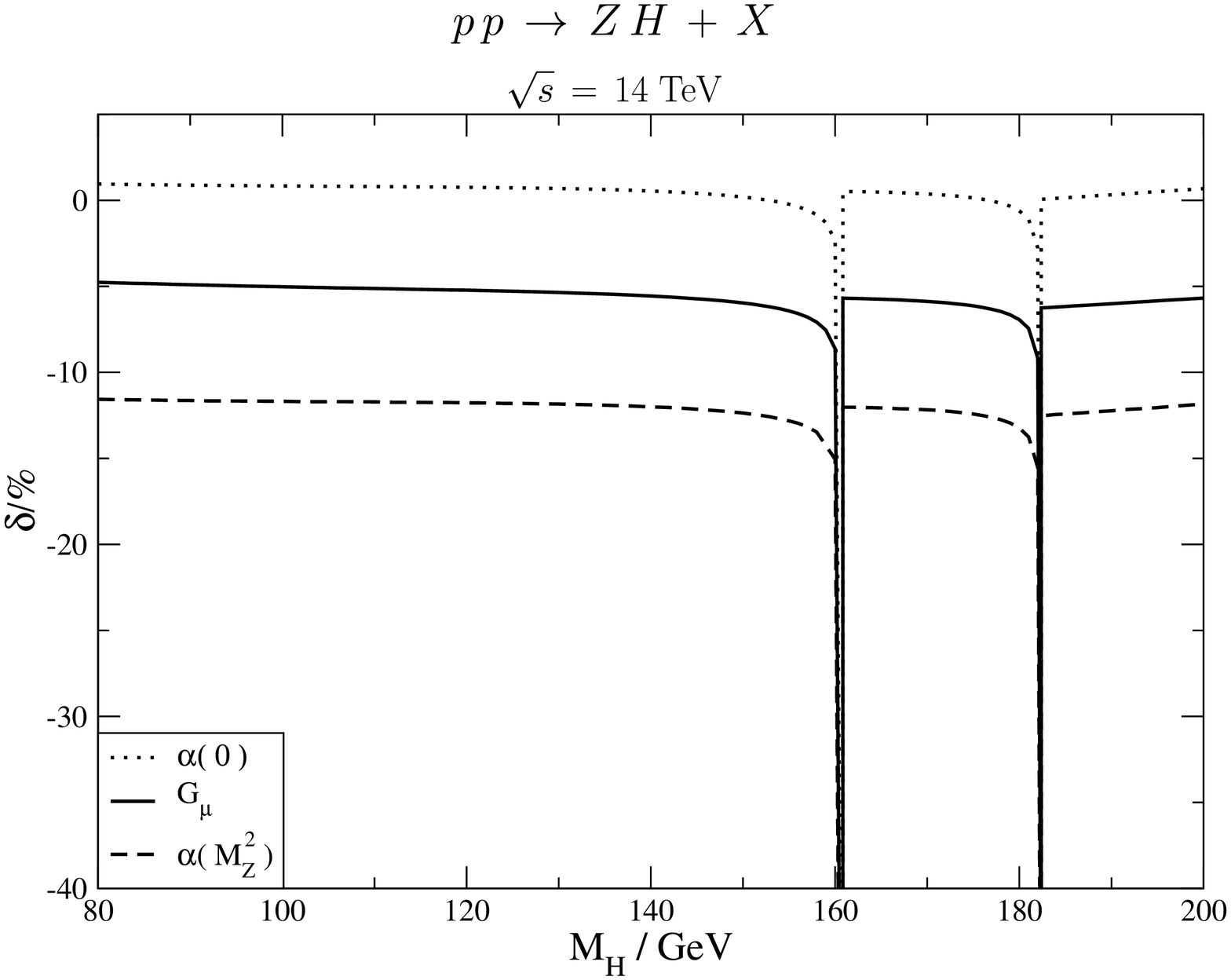,%
        bbllx=35pt,bblly=50pt,bburx=719pt,bbury=582pt,scale=0.5}
\vspace*{5mm}
\caption{\label{fig:lhczh}Relative electroweak correction $\delta$ as 
 a function of the Higgs-boson mass for the total cross section $pp\to
 ZH + X$ ($\sqrt{s}=14\;\mathrm{TeV}$).}
\end{center}
\end{figure}

Figures~\ref{fig:lhcwh} and \ref{fig:lhczh} and \reftas{tab:lhcwh} and 
\ref{tab:lhczh} show the corresponding results for $p p \to W^+ H + X$
and $p p \to ZH + X$ at the LHC. The corrections are similar in size
to those at the Tevatron and reduce the cross section by 5--10\% in
the $\GF$-scheme and by 12--17\% in the $\alpha(\MZ^2)$-scheme.
We note that the electroweak corrections to $p p \to W^- H + X$ at the
LHC differ from those to $p p \to W^+ H + X$ by less than about $2\%$.

\begin{figure}[p]
\begin{center}

\vspace*{-5mm}

\epsfig{file=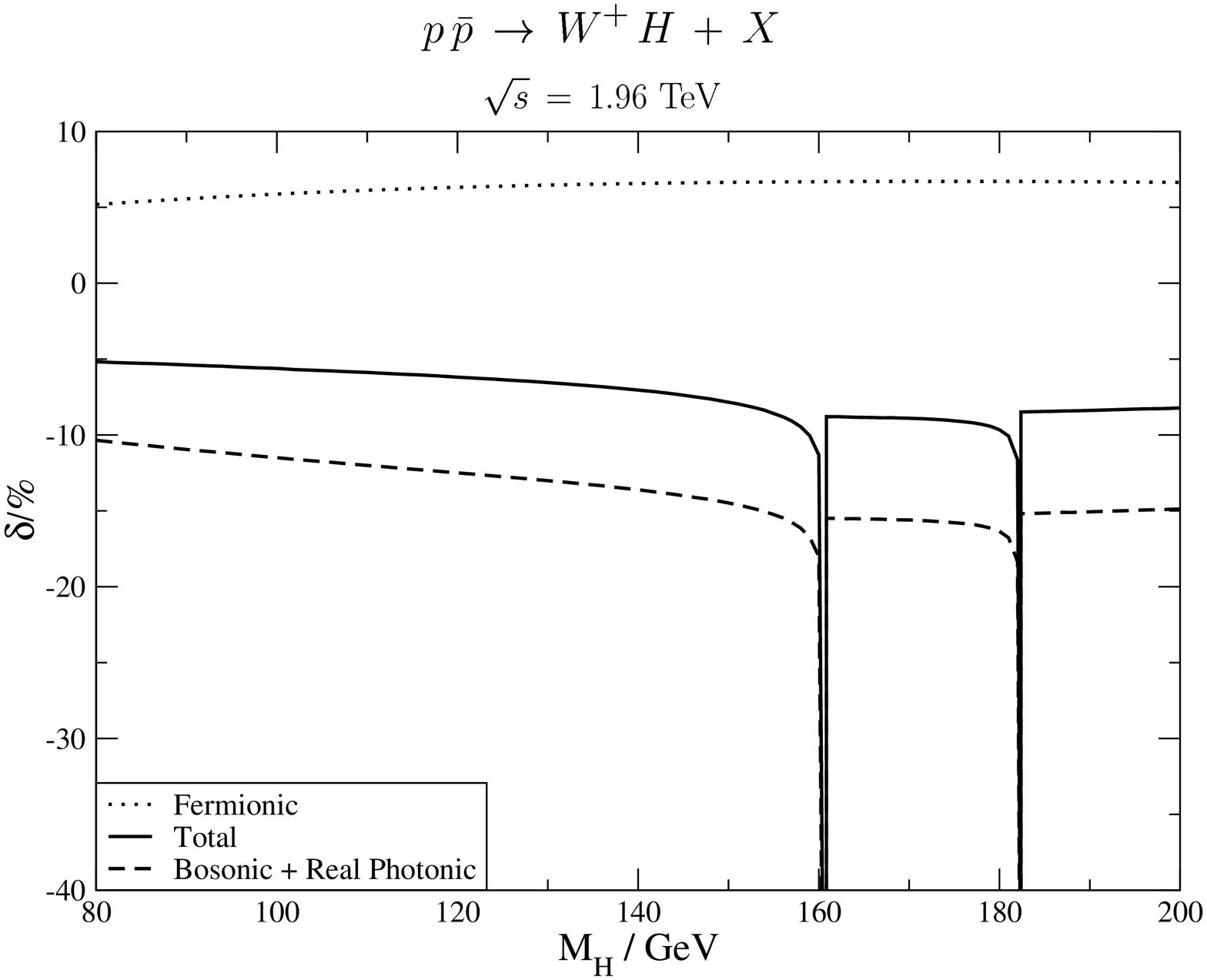,%
        bbllx=35pt,bblly=50pt,bburx=719pt,bbury=582pt,scale=0.5}
\vspace*{5mm}
\caption{\label{fig:indwh}Different contributions to the relative
 electroweak correction $\delta$ in the $\GF$-scheme as a function of
 the Higgs-boson mass for the total cross section $p\bar{p}\to W^+H +
 X$ ($\sqrt{s}=1.96\;\mathrm{TeV}$).}

\vspace*{5mm}

\epsfig{file=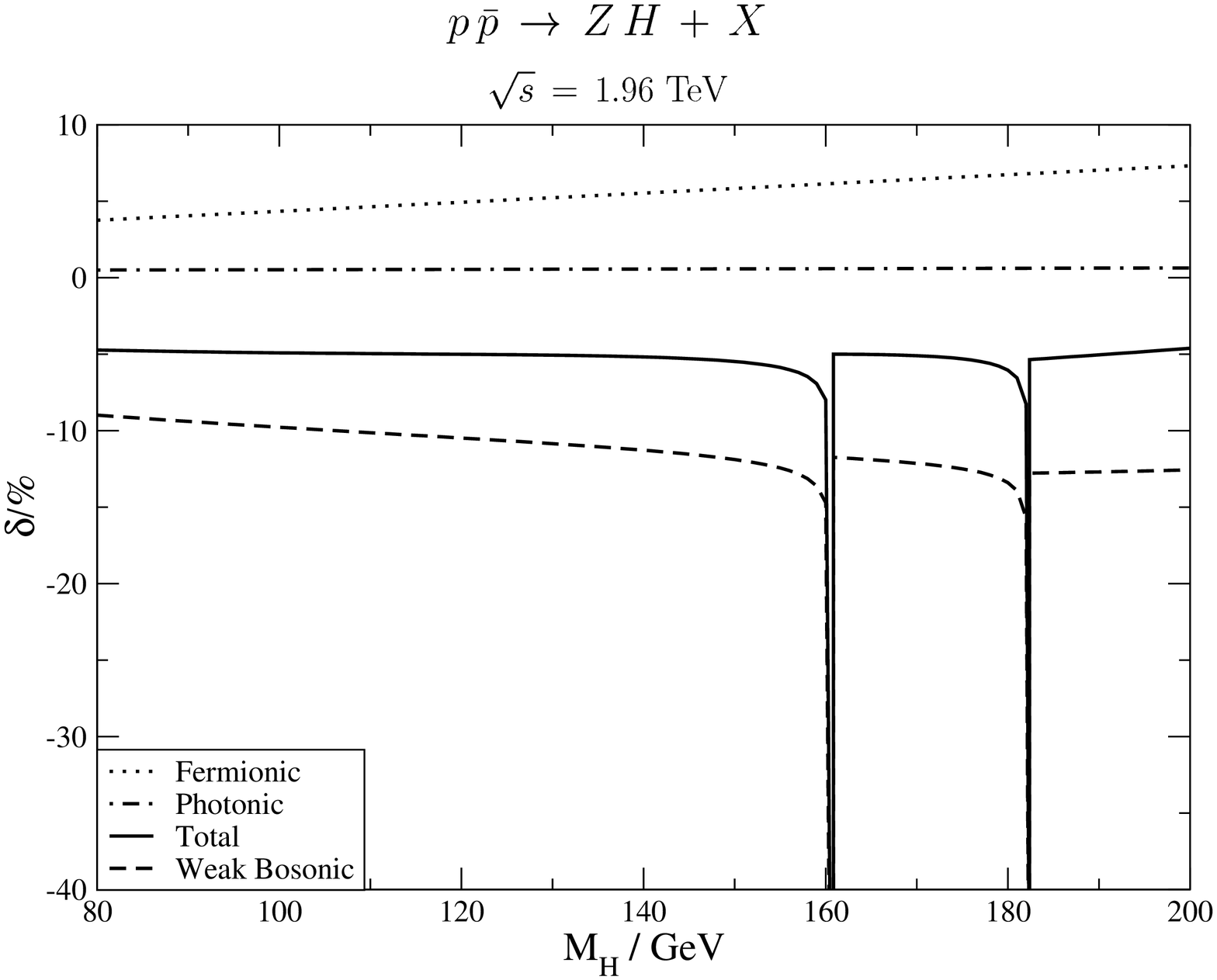,%
        bbllx=35pt,bblly=50pt,bburx=719pt,bbury=582pt,scale=0.5}
\vspace*{5mm}
\caption{\label{fig:indzh}Different contributions to the relative 
 electroweak correction $\delta$ in the $\GF$-scheme as a function of
 the Higgs-boson mass for the total cross section $p\bar{p}\to Z H +
 X$ ($\sqrt{s}=1.96\;\mathrm{TeV}$).}
\end{center}
\end{figure}

\begin{figure}[p]
\begin{center}

\vspace*{-5mm}

\epsfig{file=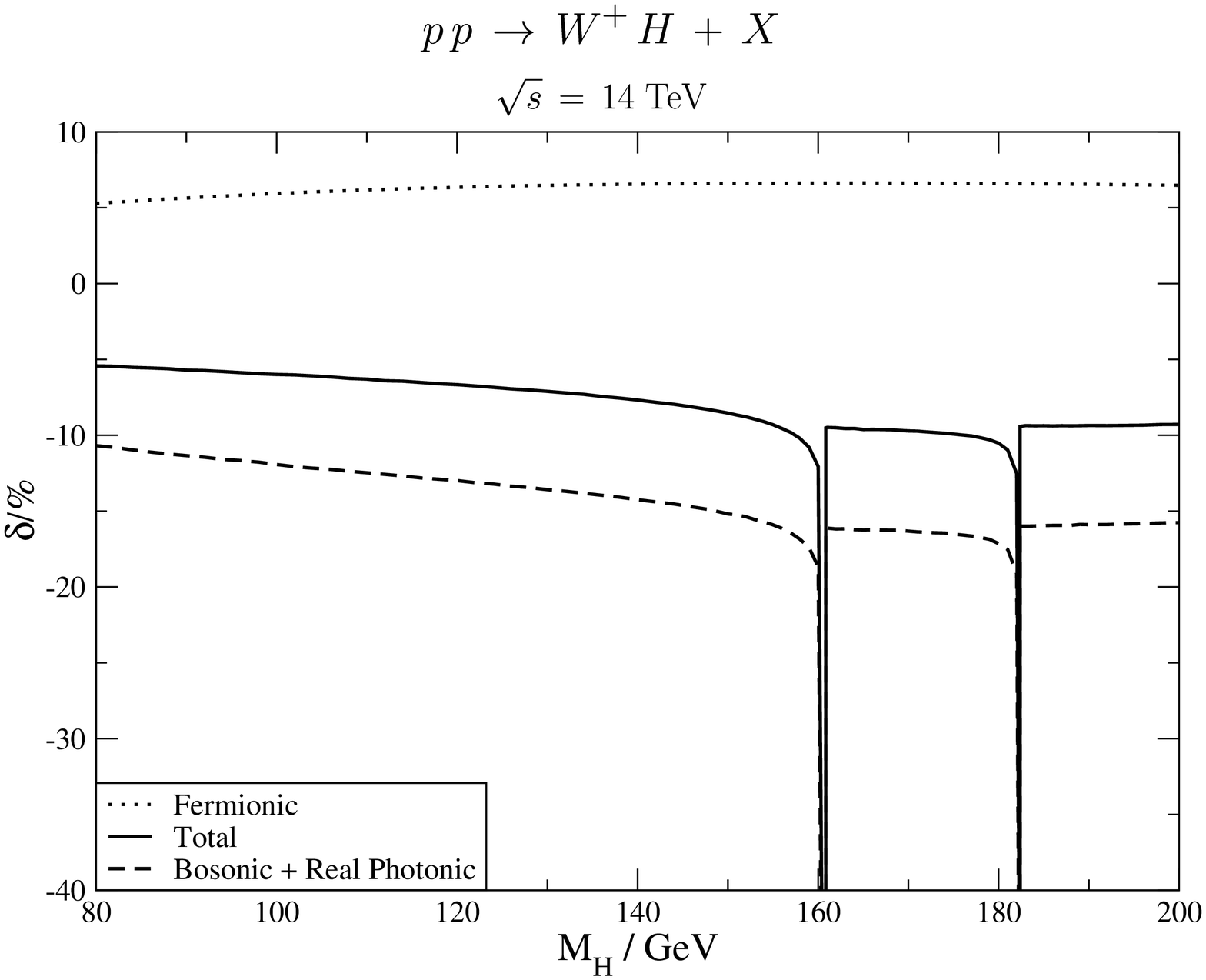,%
        bbllx=35pt,bblly=50pt,bburx=719pt,bbury=582pt,scale=0.5}
\vspace*{5mm}
\caption{\label{fig:indwhlhc}Different contributions to the relative
 electroweak correction $\delta$ in the $\GF$-scheme as a function of
 the Higgs-boson mass for the total cross section $p p\to W^+ H + X$
 ($\sqrt{s}=14\;\mathrm{TeV}$).}

\vspace*{5mm}

\epsfig{file=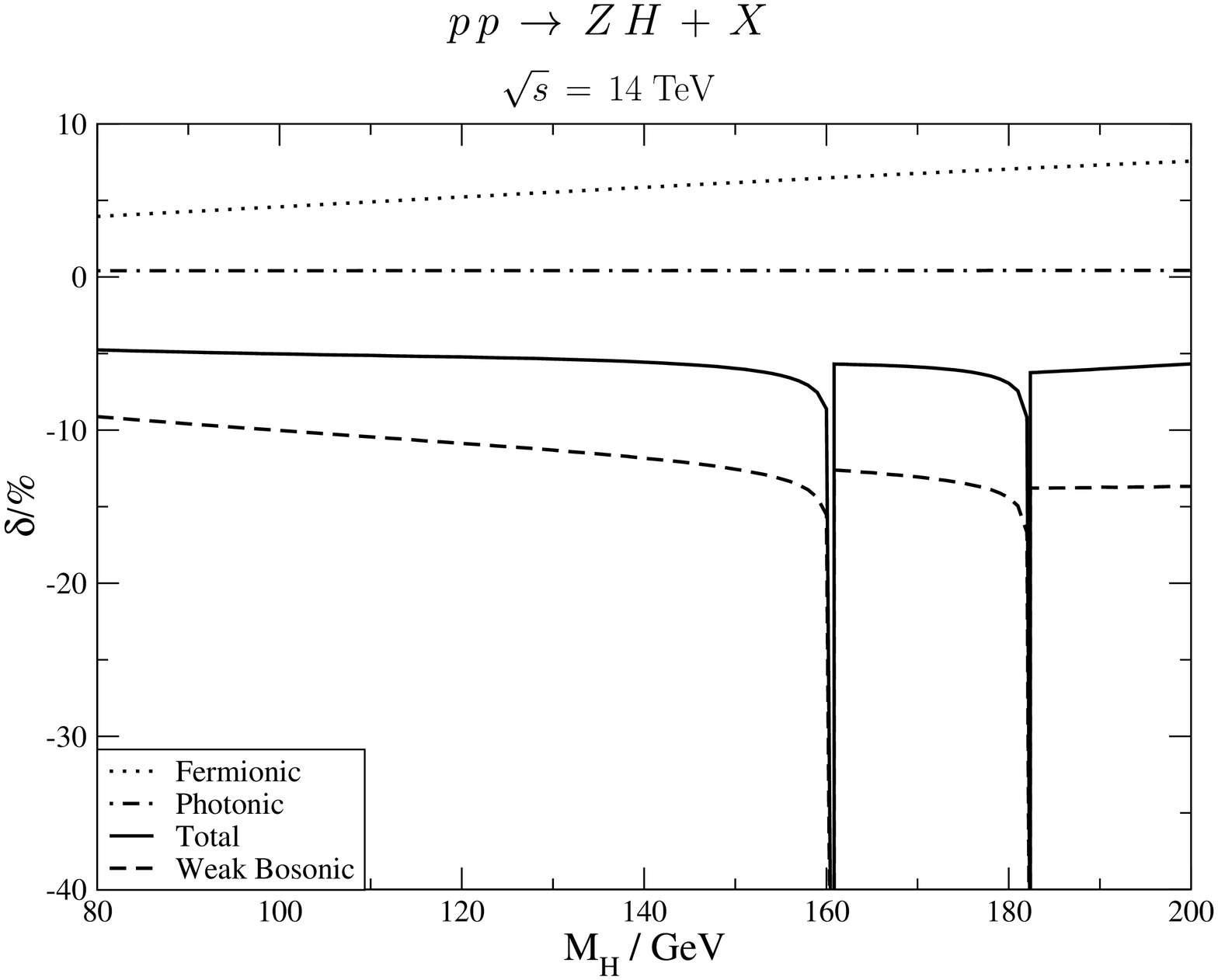,%
        bbllx=35pt,bblly=50pt,bburx=719pt,bbury=582pt,scale=0.5}
\vspace*{5mm}
\caption{\label{fig:indzhlhc}Different contributions to the relative
 electroweak correction $\delta$ in the $\GF$-scheme as a function of
 the Higgs-boson mass for the total cross section $p p\to Z H + X$
 ($\sqrt{s}=14\;\mathrm{TeV}$).}
\end{center}
\end{figure}

In order to unravel the origin of the electroweak corrections we
display the contributions of individual gauge-invariant building
blocks. Figure~\ref{fig:indwh} separates the fermionic corrections
(comprising all diagrams with closed fermion loops) from the remaining
bosonic contributions to $p\bar p \to W^+ H + X$ at the Tevatron in
the $\GF$-scheme. We observe that the bosonic corrections are dominant
and that bosonic and fermionic contributions partly compensate each
other. A similar result is found for the $p\bar p \to Z H + X$ cross
section, where we display the gauge-invariant contributions from
(photonic) QED corrections, fermionic corrections, and weak bosonic
corrections in \reffi{fig:indzh}. Note that large logarithmic
corrections from initial-state photon radiation have been absorbed
into the quark distribution functions. The remainder of the QED
corrections turns out to be strongly suppressed with respect to the
fermionic and weak bosonic corrections. A similar pattern is observed
for the $p p \to W^+ H + X$ and $p p \to Z H + X$ cross sections at
the LHC, see \reffis{fig:indwhlhc} and \ref{fig:indzhlhc}.

At first sight, the large size of the non-universal corrections, i.e.\
corrections that are not due to the running of $\alpha(k^2)$, photon
radiation, or other universal effects, might be surprising. However, a
similar pattern has already been observed in the electroweak
corrections to the processes $\Pep\Pem\to ZH$ \cite{Fleischer:1982af}
and $\Pep\Pem\to Z^*H\to\nu\bar\nu H$
\cite{Belanger:2002me,Denner:2003iy}. Also there 
large non-universal fermionic and bosonic corrections of opposite sign
occur. It was also observed that the corrections cannot be
approximated by simple formulae resulting from appropriate asymptotic
limits.%
\footnote{In \citere{Li:1998wy} the part of the fermion-loop correction
that is enhanced by an explicit factor $\alpha\Mt^2/\MW^2$ was calculated.
Moreover, in the second paper of \citere{Li:1998wy} also diagrams with 
internal Higgs bosons were taken into account.
Using $\alpha=1/128$, which roughly corresponds to the $\alpha(\MZ^2)$-scheme,
these authors find about $-1\%$ to $-2\%$ for the sum of these corrections, 
which were assumed to be the leading ones.
This has to be compared with our result of about $-12\%$ for the full
${\cal O}(\alpha)$ corrections in the $\alpha(\MZ^2)$-scheme.}
For instance, taking the large top-mass limit ($\Mt\to\infty$)
in the fermionic corrections to $WH$ production in the $\GF$-scheme
(see \refse{se:ren_ips}), the leading term in the relative correction
is given by $\delta^{\mathrm{top}}_{q\bar q'\to WH}
\Big|_{\GF}\approx-1.6\%$, which even differs in sign from the full
result (see \reffi{fig:indwh}). The reason for this failure is that
the relevant scale in the $WWH$ vertex, from which the leading $\Mt^2$
term in the limit $\Mt\to\infty$ results, is set by the variable $\hat
s$ which is not much smaller than but rather of the same order as 
$\Mt^2$.  For the $ZH$ channel, we get $\delta^{\mathrm{top}}_{u\bar
u\to ZH} \Big|_{\GF}
\approx \delta^{\mathrm{top}}_{d\bar d\to ZH} \Big|_{\GF} \approx  
- 1\%$, again reflecting the failure of the heavy-top limit as a
suitable approximation.  Concerning the weak bosonic corrections,
large negative contributions are expected in the high-energy limit
owing to the occurrence of Sudakov logarithms of the form
$-\alpha/\pi\log^2(\hat s/\MW^2)$.  However, the relevant partonic CM
energies $\sqrt{\hat s}$ are not yet large enough for the Sudakov
logarithms to provide a good approximation for the full corrections.

\subsection{The cross section at NLO}
\begin{figure}[p]
\begin{center}

\vspace*{-8mm}

\epsfig{file=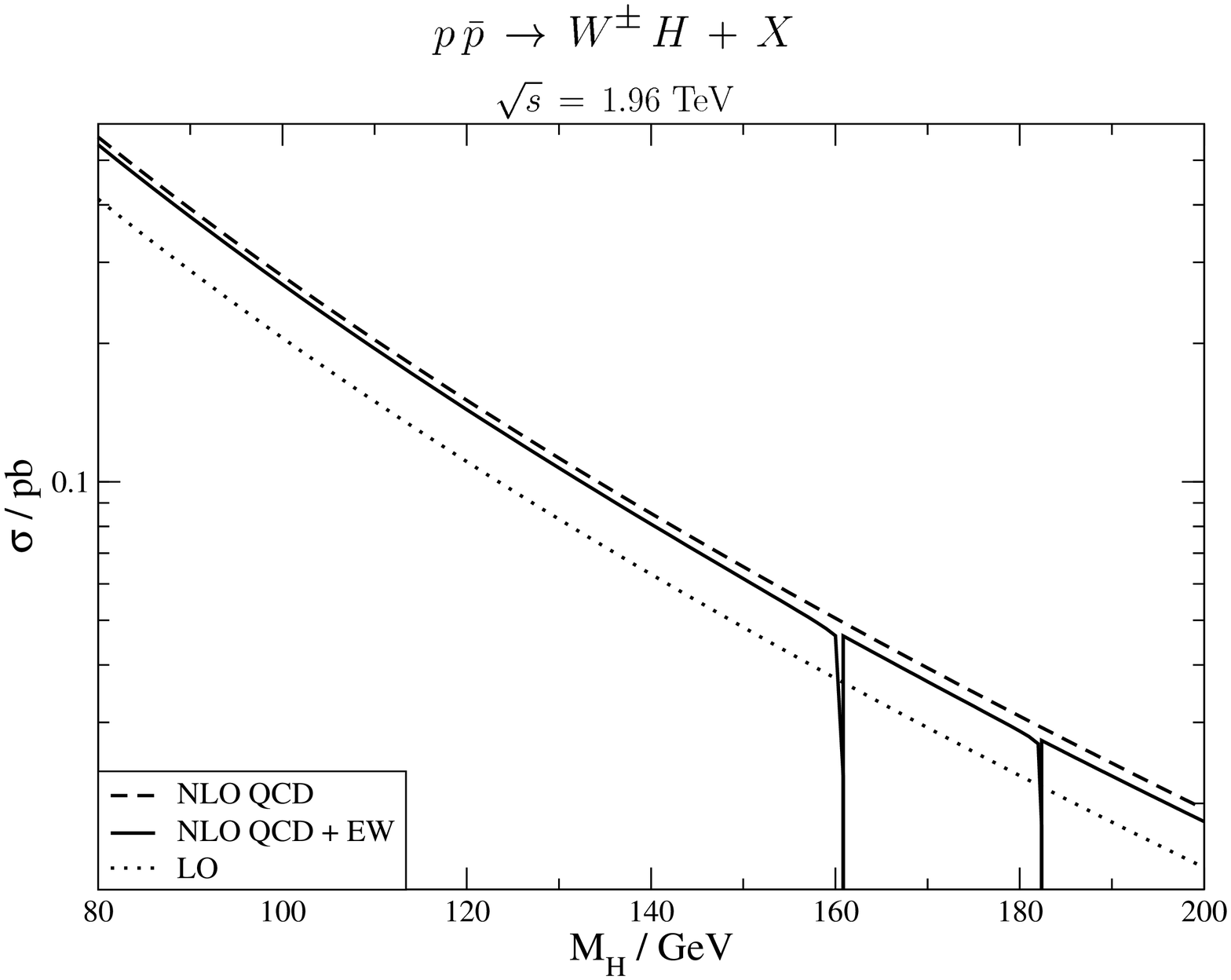,%
        bbllx=35pt,bblly=50pt,bburx=719pt,bbury=582pt,scale=0.5}
\vspace*{6mm}
\caption{\label{fig:totwhtev}Total cross section for 
 $p\bar{p}\to W^{\pm} H + X$ (sum of $W^+H$ and $W^-H$) at the
 Tevatron ($\sqrt{s}=1.96\;\mathrm{TeV}$) in LO, NLO QCD, and
 including NLO QCD and electroweak corrections in the
 $\GF$-scheme. The renormalization and factorization scales have been
 set to the invariant mass of the Higgs--vector-boson pair, $\mu =
 \sqrt{s_{VH}}$. CTEQ6L1 and CTEQ6M \cite{Pumplin:2002vw} parton
 distribution functions have been adopted at LO and ${\cal
 O}(\alpha_{\rm s})$, respectively.}

\vspace*{0mm}


\epsfig{file=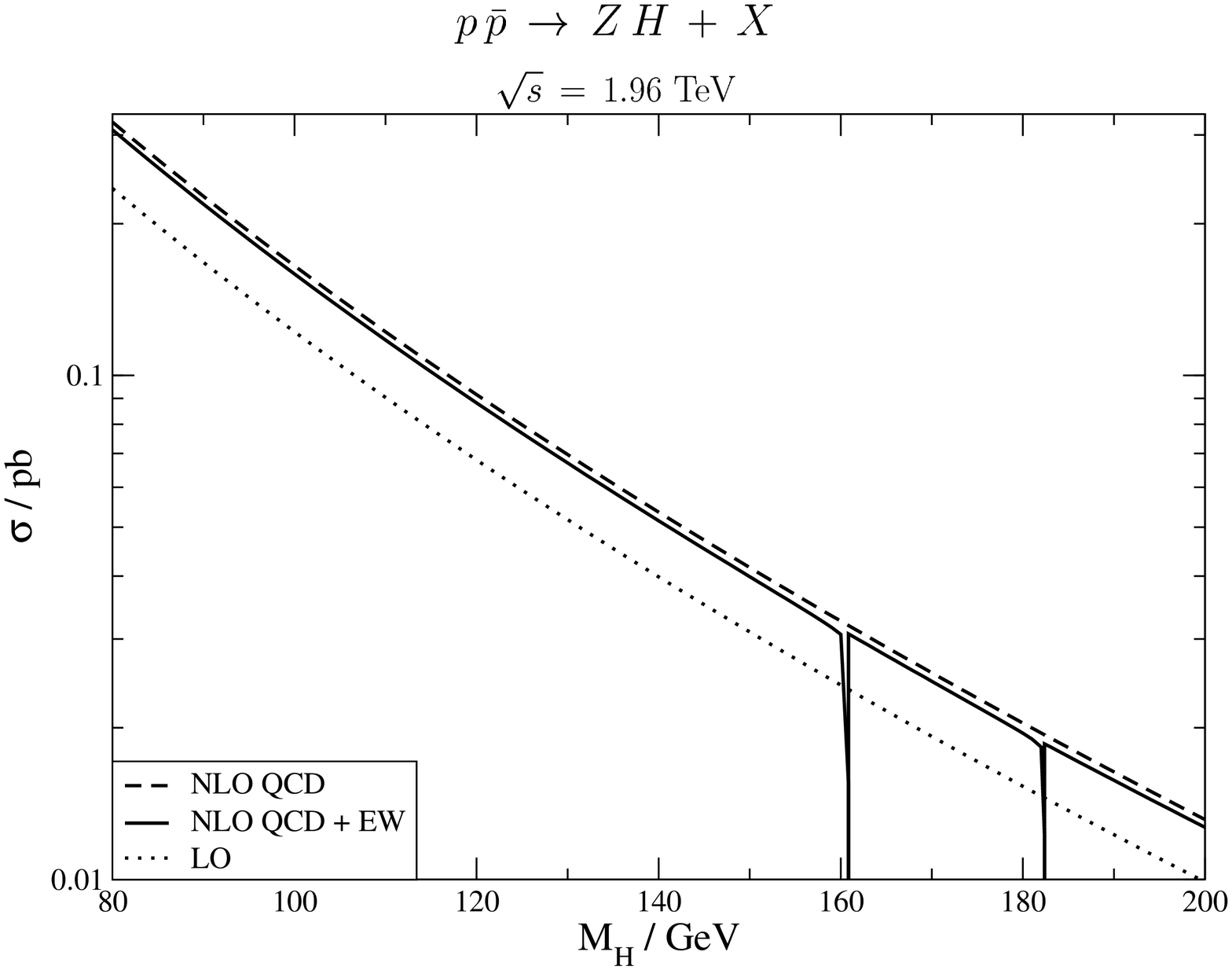,%
        bbllx=35pt,bblly=50pt,bburx=719pt,bbury=582pt,scale=0.5}
\vspace*{6mm}
\caption{\label{fig:totzhtev}Total cross section for 
 $p\bar{p}\to Z H + X$ at the Tevatron ($\sqrt{s}=1.96\;\mathrm{TeV}$)
 in LO, NLO QCD, and including NLO QCD and electroweak
 corrections in the $\GF$-scheme. 
}
\vspace*{5mm}

\end{center}
\end{figure}

\begin{figure}[p]
\begin{center}

\vspace*{-6mm}

\epsfig{file=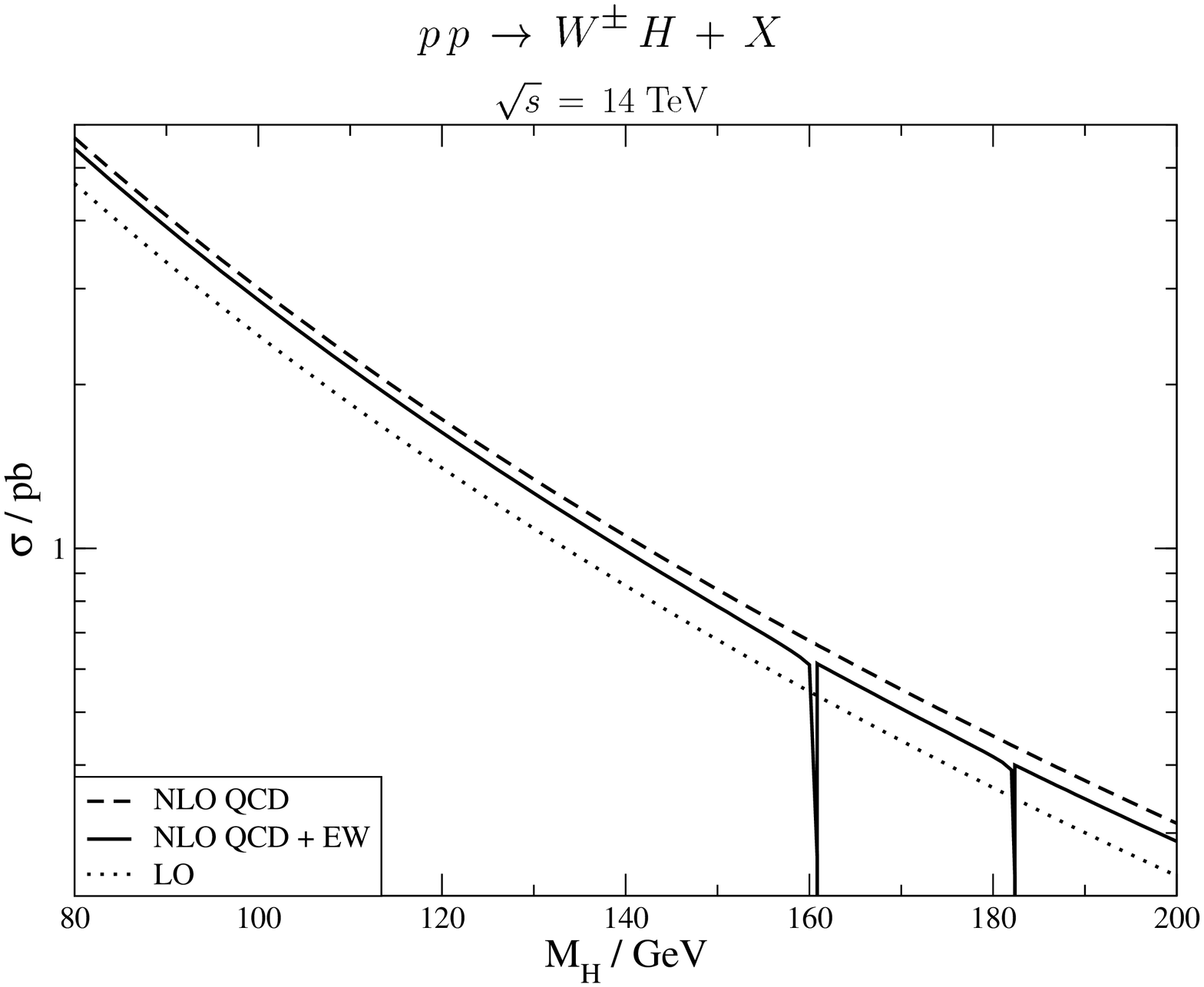,%
        bbllx=35pt,bblly=50pt,bburx=719pt,bbury=582pt,scale=0.5}
\vspace*{6mm}
\caption{\label{fig:totwhlhc}Total cross section for 
 $pp\to W^{\pm} H + X$ at the LHC ($\sqrt{s}=14\;\mathrm{TeV}$)
 in LO, NLO QCD, and including NLO QCD and electroweak
 corrections in the $\GF$-scheme. 
}

\vspace*{5mm}


\epsfig{file=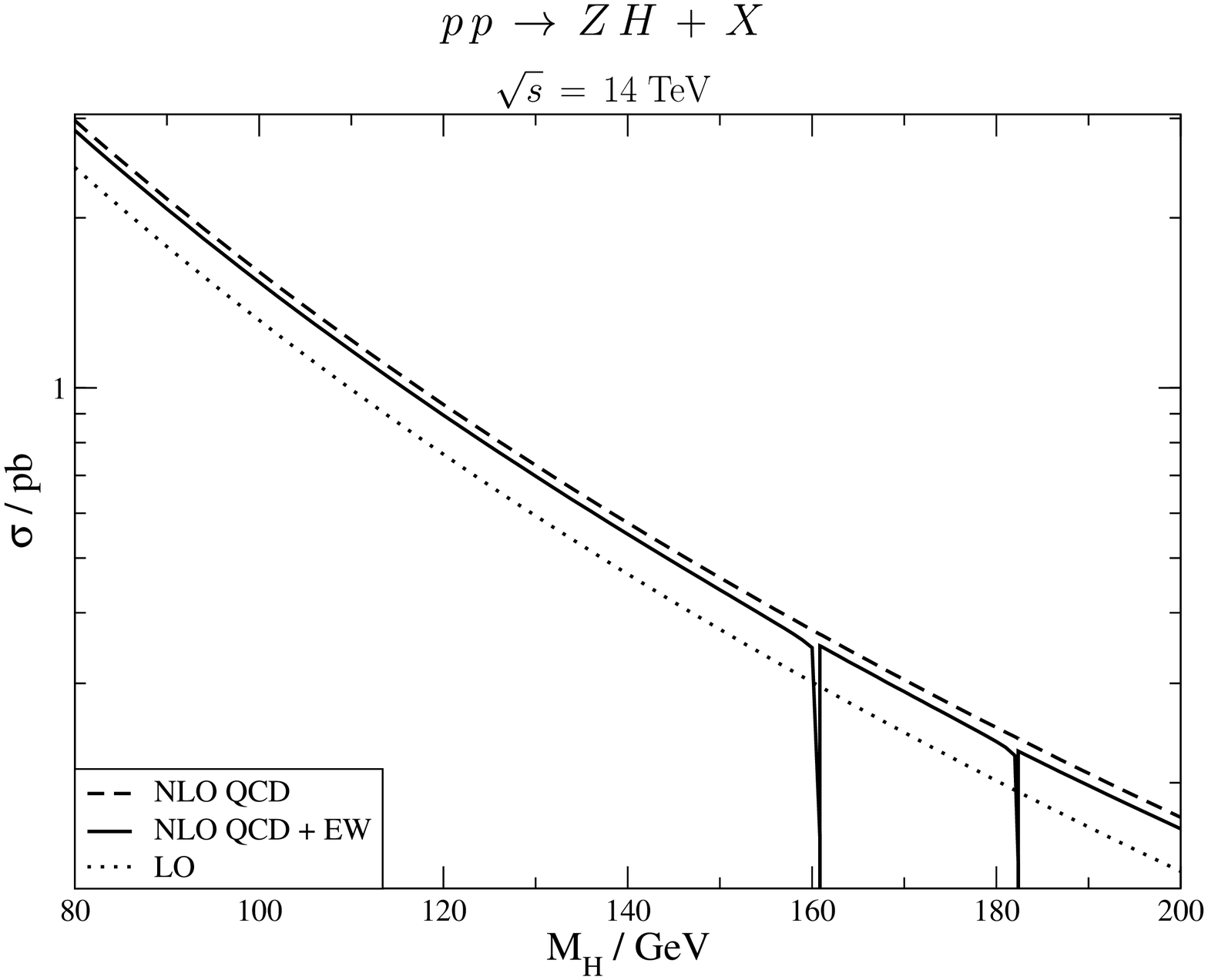,%
        bbllx=35pt,bblly=50pt,bburx=719pt,bbury=582pt,scale=0.5}
\vspace*{6mm}
\caption{\label{fig:totzhlhc}Total cross section for 
 $pp \to Z H + X$ at the LHC ($\sqrt{s}=14\;\mathrm{TeV}$)
 in LO, NLO QCD, and including NLO QCD and electroweak
 corrections in the $\GF$-scheme. 
}
\end{center}
\end{figure}

In this subsection we present the cross section prediction for
associated $WH$ and $ZH$ production at the Tevatron and at the LHC,
including the NLO order electroweak and QCD corrections, and we
quantify the residual theoretical uncertainty due to scale variation
and the parton distribution functions. The total cross sections for
the processes $p\bar p/pp \to W^{\pm}H+X$ (sum of $W^+H$ and $W^-H$)
and $p\bar{p}/pp \to ZH+X$ at the Tevatron and the LHC are displayed
in \reffis{fig:totwhtev}--\ref{fig:totzhlhc}. Representative results
are listed in \reftas{tab:totwhtev}--\ref{tab:totzhlhc}. For the
central renormalization and factorization scale $\mu = \mu_0 =
\sqrt{s_{VH}}$ the NLO QCD corrections increase the LO cross section
by typically 20--25\%. As discussed in detail in \refse{sec:ewk}, the
NLO electroweak corrections are sizeable and decrease the cross
section by 5--10\% in the $\GF$-scheme. The size of the ${\cal
O}(\alpha_{\rm s})$ and ${\cal O}(\alpha)$ corrections does not
depend strongly on the Higgs-boson mass.

The NLO prediction is very stable under variation of the QCD
renormalization and factorization scales. We have varied both scales
independently in the range $\mu_0/5 < \mu < 5\mu_0$.  For both the
Tevatron and the LHC, the cross section increases monotonically with
decreasing renormalization scale. At the Tevatron, the maximal
(minimal) cross section is obtained choosing both the renormalization
and factorization scales small (large). At the LHC, in contrast, the
maximal (minimal) cross section corresponds to choosing a large
(small) factorization scale. From the numbers listed in
\reftas{tab:totwhtev}--\ref{tab:totzhlhc} one can conclude that
the theoretical uncertainty introduced by varying the QCD scales in
the range $\mu_0/5 < \mu < 5\mu_0$ is less than approximately
$10\%$. We have verified that the QED factorization-scale dependence
of the ${\cal O}(\alpha)$-corrected cross section is below $1\%$ and
thus negligible compared to the other theoretical uncertainties. The
QED scale dependence should be reduced further when using QED-improved
parton densities.

We have also studied the uncertainty in the cross-section prediction
due to the error in the parametrization of the parton densities. To
this end we have compared the NLO cross section evaluated using the
default CTEQ6~\cite{Pumplin:2002vw} parametrization with the cross
section evaluated using the MRST2001~\cite{Martin:2002aw}
parametrization. The results are collected in
\reftas{tab:pdfwhtev}--\ref{tab:pdfzhlhc}. Both the CTEQ and MRST 
parametrizations include parton-distribution-error packages which
provide a quantitative estimate of the corresponding uncertainties in
the cross sections.%
\footnote{In addition, the MRST~\cite{Martin:2001es} parametrization allows to
study the uncertainty of the NLO cross section due to the variation of
$\alpha_{\rm s}$. For associated $WH$ and $ZH$ hadroproduction, the
sensitivity of the theoretical prediction to the variation of
$\alpha_{\rm s}$ ($\alpha_{\rm s}(\MZ^2) = 0.119\pm 0.02$) turns out
to be below $2\%$.}  Using the parton-distribution-error packages and
comparing the CTEQ and MRST2001 parametrizations, we find that the
uncertainty in predicting the processes $p\bar p/pp \to W^{\pm}H+X$
and $p\bar{p}/pp \to ZH+X$ at the Tevatron and the LHC due to the
parametrization of the parton densities is less than approximately
$5\%$.

\section{Conclusions}
\label{se:concl}

We have calculated the electroweak ${\cal O}(\alpha)$ corrections to
Higgs-boson production in association with $W$ or $Z$~bosons at hadron
colliders. These corrections decrease the theoretical prediction by up
to 5--10\%, depending in detail on the Higgs-boson mass and the 
input-parameter scheme. We have updated the cross section prediction for
associated $WH$ and $ZH$ production at the Tevatron and at the LHC,
including the next-to-leading order electroweak and QCD corrections.
Finally, the remaining theoretical uncertainty has been studied by
varying the renormalization and factorization scales and by taking into
account the uncertainties in the parton distribution functions.
We find that the scale dependence is reduced to about $10\%$ at
next-to-leading order, while the uncertainty due to the parton
densities is less than about $5\%$.

\section*{Acknowledgement}
This work has been supported in part by the European Union under
contract HPRN-CT-2000-00149. M.~L.~Ciccolini is partially supported by
ORS Award ORS/2001014035.


\begin{table}[p]
\centerline{
\begin{tabular}{|c||c|c||c|c||c|c|}
\hline
$M_H/\mathrm{GeV}$ &
$\sigma_0|_{\alpha(0)}/\mathrm{pb} $ & 
$\delta|_{\alpha(0)}/\% $ & 
$\sigma_0|_{\alpha(M_Z^2)}/\mathrm{pb} $ & 
$\delta|_{\alpha(M_Z^2)}/\% $ & 
$\sigma_0|_{G_\mu}/\mathrm{pb} $ & 
$\delta|_{G_\mu}/\% $ \\ \hline \hline
80.00	& 0.1926(1)   &   0.57(1)  & 0.2175(1)   & $-11.99(1)$ & 0.2058(1)   & $-5.17(1)$ \\ \hline
100.00	& 0.09614(1)  &   0.25(1)  & 0.1086(1)   & $-12.29(1)$ & 0.1028(1)   & $-5.61(1)$ \\ \hline
120.00	& 0.05176(1)  & $-0.17(1)$ & 0.05846(1)  & $-12.75(1)$ & 0.05532(1)  & $-6.19(1)$ \\ \hline
140.00	& 0.02945(1)  & $-0.89(1)$ & 0.03327(1)  & $-13.53(1)$ & 0.03149(1)  & $-7.05(1)$ \\ \hline
170.00	& 0.01367(1)  & $-2.55(1)$ & 0.01544(1)  & $-15.30(1)$ & 0.01461(1)  & $-8.90(1)$ \\ \hline
190.00	& 0.008517(1) & $-1.97(1)$ & 0.009624(1) & $-14.66(1)$ & 0.009106(1) & $-8.38(1)$ \\ \hline
\end{tabular}}
\caption{Total lowest-order hadronic cross section 
 $\sigma_0(p \bar p \;\to\; W^+\,H+X)$ and corresponding relative
 electroweak correction $\delta$ ($\sqrt{s}=1.96\;\mathrm{TeV}$).
 Results are presented for the $\alpha(0)$-, $\alpha(\MZ^2)$-, and
 $\GF$-schemes. The integration error is given in brackets.}
\label{tab:tevwh} 

\vspace*{1cm}

\centerline{
\begin{tabular}{|c||c|c||c|c||c|c|}
\hline
$M_H/\mathrm{GeV}$ & $\sigma_0|_{\alpha(0)}/\mathrm{pb} $
& $\delta|_{\alpha(0)}/\% $ &
$\sigma_0|_{\alpha(M_Z^2)}/\mathrm{pb} $ &
$\delta|_{\alpha(M_Z^2)}/\% $ & $\sigma_0|_{G_\mu}/\mathrm{pb} $ &
$\delta|_{G_\mu}/\% $ \\ \hline \hline
80.00  & 0.2199(1)   & 0.99(1) & 0.2484(1)  & $-11.52(1)$ & 0.2350(1)   & $-4.73(1)$ \\ \hline
100.00 & 0.1142(1)   & 0.95(1) & 0.1290(1)  & $-11.57(1)$ & 0.1221(1)   & $-4.91(1)$ \\ \hline
120.00 & 0.06358(1)  & 0.97(1) & 0.07182(1) & $-11.55(1)$ & 0.06796(1)  & $-5.01(1)$ \\ \hline
140.00 & 0.03727(1)  & 0.91(1) & 0.04211(1) & $-11.61(1)$ & 0.03984(1)  & $-5.18(1)$ \\ \hline
170.00 & 0.01799(1)  & 1.12(1) & 0.02032(1) & $-11.38(1)$ & 0.01922(1)  & $-5.10(1)$ \\ \hline
190.00 & 0.01148(1)  & 1.26(1) & 0.01297(1) & $-11.24(1)$ & 0.01227(1)  & $-5.04(1)$ \\ \hline
\end{tabular}}
\caption{Total lowest-order hadronic cross section 
$\sigma_0(p \bar p \;\to\; Z\,H+X)$ and corresponding relative
electroweak correction $\delta$ ($\sqrt{s}=1.96\;\mathrm{TeV}$).}
\label{tab:tevzh} 
\end{table}


\begin{table}[p]

\centerline{
\begin{tabular}{|c||c|c||c|c||c|c|}
\hline
$M_H/\mathrm{GeV}$ &
$\sigma_0|_{\alpha(0)}/\mathrm{pb} $ & 
$\delta|_{\alpha(0)}/\% $ & 
$\sigma_0|_{\alpha(M_Z^2)}/\mathrm{pb} $ & 
$\delta|_{\alpha(M_Z^2)}/\% $ & 
$\sigma_0|_{G_\mu}/\mathrm{pb} $ & 
$\delta|_{G_\mu}/\% $ \\ \hline \hline
80.00	& 2.660(1)  &   0.31(1)  & 3.005(1)  & $-12.22(2)$ & 2.844(1)  & $-5.43(1)$ \\ \hline 
100.00	& 1.410(1)  & $-0.11(1)$ & 1.594(1)  & $-12.67(2)$ & 1.508(1)  & $-5.99(1)$ \\ \hline 
120.00	& 0.8114(2) & $-0.65(1)$ & 0.9166(2) & $-13.24(2)$ & 0.8673(2) & $-6.67(1)$ \\ \hline 
140.00	& 0.4967(1) & $-1.49(1)$ & 0.5610(1) & $-14.16(2)$ & 0.5309(1) & $-7.68(1)$ \\ \hline 
170.00	& 0.2605(1) & $-3.33(1)$ & 0.2942(1) & $-16.12(2)$ & 0.2784(1) & $-9.72(2)$ \\ \hline 
190.00	& 0.1776(1) & $-2.92(1)$ & 0.2007(1) & $-15.67(2)$ & 0.1899(1) & $-9.36(1)$ \\ \hline 
\end{tabular}}
\caption{Total lowest-order hadronic cross section 
$\sigma_0(p p \;\to\; W^+\,H+X)$ and corresponding relative
electroweak correction $\delta$ ($\sqrt{s}=14\;\mathrm{TeV}$).}
\label{tab:lhcwh} 

\vspace{1 cm}

\centerline{
\begin{tabular}{|c||c|c||c|c||c|c|}
\hline
$M_H/\mathrm{GeV}$ &
$\sigma_0|_{\alpha(0)}/\mathrm{pb} $ & 
$\delta|_{\alpha(0)}/\% $ & 
$\sigma_0|_{\alpha(M_Z^2)}/\mathrm{pb} $ & 
$\delta|_{\alpha(M_Z^2)}/\% $ & 
$\sigma_0|_{G_\mu}/\mathrm{pb} $ & 
$\delta|_{G_\mu}/\% $ \\ \hline \hline
80.00	& 2.299(1)  & 0.95(1) & 2.595(1)  & $-11.56(1)$ & 2.457(1)  & $-4.77(1)$ \\ \hline
100.00	& 1.232(1)  & 0.83(1) & 1.392(1)  & $-11.68(1)$ & 1.317(1)  & $-5.03(1)$ \\ \hline
120.00	& 0.7134(1) & 0.76(1) & 0.8058(1) & $-11.77(1)$ & 0.7630(1) & $-5.22(1)$ \\ \hline
140.00	& 0.4381(1) & 0.54(1) & 0.4950(1) & $-12.01(1)$ & 0.4684(1) & $-5.56(1)$ \\ \hline
170.00	& 0.2297(1) & 0.37(1) & 0.2595(1) & $-12.18(1)$ & 0.2456(1) & $-5.88(1)$ \\ \hline
190.00	& 0.1563(1) & 0.32(1) & 0.1765(1) & $-12.23(1)$ & 0.1670(1) & $-6.01(1)$ \\ \hline
\end{tabular}}
\caption{Total lowest-order hadronic cross section 
$\sigma_0(p p \;\to\; Z\,H+X)$ and corresponding relative
electroweak correction $\delta$ ($\sqrt{s}=14\;\mathrm{TeV}$).}
\label{tab:lhczh} 

\end{table}


\begin{table}[p]

\centerline{
\begin{tabular}{|c||c|c|c||c|c|}
\hline
\raisebox{-1.5em}[0em][0em]{$M_H/\mathrm{GeV}$} &
\raisebox{-1.5em}[0em][0em]{$\sigma_0|_{G_\mu}/\mathrm{pb} $} & 
\raisebox{-1.5em}[0em][0em]{$\sigma^{\rm QCD}_{\rm NLO}/\mathrm{pb} $} & 
\raisebox{-1.5em}[0em][0em]{$\sigma^{\rm QCD+EW}_{\rm NLO}/\mathrm{pb} $} & 
\multicolumn{2}{|c|}{$\sigma^{\rm QCD+EW}_{\rm NLO}/\mathrm{pb}\rule[-3mm]{0mm}{9mm}  $}
\\ \cline{5-6}
 &
 & 
 & 
 & 
$\mu_R = 5\mu_0$ & $\mu_R = \mu_0/5$\\[-2mm]
 &
 & 
 & 
 & 
$\mu_F = 5\mu_0$ & $\mu_F = \mu_0/5$\\[0.5mm]
 \hline \hline
80.00	& 0.4117(1)	& 0.5616(2)	& 0.5404(2)	& 0.5033(1)	& 0.5838(1)  \\ \hline
100.00	& 0.2056(1)	& 0.2801(1)	& 0.2685(1)	& 0.2482(1)	& 0.2911(1)  \\ \hline
120.00	& 0.1106(1)	& 0.1504(1)	& 0.1436(1)	& 0.1318(1)	& 0.1562(1)  \\ \hline
140.00	& 0.06297(1)	& 0.08536(1)	& 0.08092(1)	& 0.07377(1)	& 0.08833(1) \\ \hline
170.00	& 0.02921(1)	& 0.03940(1)	& 0.03679(1)	& 0.03318(1)	& 0.04037(1) \\ \hline
190.00	& 0.01821(1)	& 0.02446(1)	& 0.02294(1)	& 0.02056(1)	& 0.02525(1) \\ \hline
\end{tabular}}
\caption{\label{tab:totwhtev}Total cross section for 
 $p\bar{p} \to W^{\pm} H + X$ (sum of $W^+H$ and $W^-H$) at the
 Tevatron ($\sqrt{s}=1.96\;\mathrm{TeV}$) in LO, NLO QCD, and
 including NLO QCD and electroweak corrections in the
 $\GF$-scheme. The renormalization scale ($\mu_R$) and the
 factorization scale ($\mu_F$) have been set to the invariant mass of
 the Higgs--vector-boson pair, $\mu = \mu_0 = \sqrt{s_{VH}}$. CTEQ6L1
 and CTEQ6M \cite{Pumplin:2002vw} parton distribution functions have
 been adopted at LO and ${\cal O}(\alpha_{\rm s})$, respectively. The
 last two columns show the minimal and maximal cross section
 prediction obtained from varying the QCD renormalization and
 factorization scales independently in the range $\mu_0/5 < \mu < 5
 \mu_0$.}

\vspace{1cm}

\centerline{
\begin{tabular}{|c||c|c|c||c|c|}
\hline 
\raisebox{-1.5em}[0em][0em]{$M_H/\mathrm{GeV}$} &
\raisebox{-1.5em}[0em][0em]{$\sigma_0|_{G_\mu}/\mathrm{pb} $} & 
\raisebox{-1.5em}[0em][0em]{$\sigma^{\rm QCD}_{\rm NLO}/\mathrm{pb} $} & 
\raisebox{-1.5em}[0em][0em]{$\sigma^{\rm QCD+EW}_{\rm NLO}/\mathrm{pb} $} & 
\multicolumn{2}{|c|}{$\sigma^{\rm QCD+EW}_{\rm NLO}/\mathrm{pb}\rule[-3mm]{0mm}{9mm}  $}
\\ 
\cline{5-6}
 &
 & 
 & 
 & 
$\mu_R = 5\mu_0$ & $\mu_R = \mu_0/5$\\[-2mm]
 &
 & 
 & 
 & 
$\mu_F = 5\mu_0$ & $\mu_F = \mu_0/5$\\[0.5mm]
\hline \hline
80.00	& 0.2350(1)	& 0.3181(1)	& 0.3070(1)	& 0.2858(1)	& 0.3317(1)  \\ \hline
100.00	& 0.1221(1)	& 0.1649(1)	& 0.1589(1)	& 0.1470(1)	& 0.1722(1)  \\ \hline
120.00	& 0.06796(1)	& 0.09160(1)	& 0.08820(1)	& 0.08111(2)	& 0.09575(1)  \\ \hline
140.00	& 0.03984(1)	& 0.05354(1)	& 0.05148(1)	& 0.04706(2)	& 0.05604(1) \\ \hline
170.00	& 0.01922(1)	& 0.02570(1)	& 0.02472(1)	& 0.02242(1)	& 0.02701(1) \\ \hline
190.00	& 0.01227(1)	& 0.01635(1)	& 0.01573(1)	& 0.01418(1)	& 0.01722(1) \\ \hline
\end{tabular}}
\caption{\label{tab:totzhtev}Total cross section for 
 $p\bar{p}\to Z H + X$ at the Tevatron ($\sqrt{s}=1.96\;\mathrm{TeV}$) in
 LO, NLO QCD, and including NLO QCD and electroweak corrections in the
 $\GF$-scheme. }

\end{table}


\begin{table}[p]

\centerline{
\begin{tabular}{|c||c|c|c||c|c|}
\hline
\raisebox{-1.5em}[0em][0em]{$M_H/\mathrm{GeV}$} &
\raisebox{-1.5em}[0em][0em]{$\sigma_0|_{G_\mu}/\mathrm{pb} $} & 
\raisebox{-1.5em}[0em][0em]{$\sigma^{\rm QCD}_{\rm NLO}/\mathrm{pb} $} & 
\raisebox{-1.5em}[0em][0em]{$\sigma^{\rm QCD+EW}_{\rm NLO}/\mathrm{pb} $} & 
\multicolumn{2}{|c|}{$\sigma^{\rm QCD+EW}_{\rm NLO}/\mathrm{pb}\rule[-3mm]{0mm}{9mm}  $}
\\ 
\cline{5-6}
 &
 & 
 & 
 & 
$\mu_R = 5\mu_0$ & $\mu_R = \mu_0/5$\\[-2mm]
 &
 & 
 & 
 & 
$\mu_F = \mu_0/5$ & $\mu_F = 5\mu_0$\\[0.5mm]
\hline \hline
80.00	& 4.679(2)	& 5.676(2)	& 5.423(2)	& 4.875(2)	& 5.749(5)  \\ \hline 
100.00	& 2.462(1)	& 3.005(1)	& 2.859(1)	& 2.606(2)	& 3.033(2)  \\ \hline 
120.00	& 1.405(1)	& 1.726(1)	& 1.633(1)	& 1.505(1)	& 1.731(1) \\ \hline 
140.00	& 0.8537(2)	& 1.054(1)	& 0.9892(3)	& 0.9204(3)	& 1.050(1) \\ \hline 
170.00	& 0.4434(1)	& 0.5504(1)	& 0.5078(1)	& 0.4782(1)	& 0.5388(3) \\ \hline 
190.00	& 0.3003(1)	& 0.3745(1)	& 0.3466(1)	& 0.3285(1)	& 0.3675(2) \\ \hline 
\end{tabular}}
\caption{\label{tab:totwhlhc}Total cross section for 
 $pp \to W^{\pm} H + X$ at the LHC ($\sqrt{s}=14\;\mathrm{TeV}$) in
 LO, NLO QCD, and including NLO QCD and electroweak corrections in the
 $\GF$-scheme. }

\vspace{1cm}

\centerline{
\begin{tabular}{|c||c|c|c||c|c|}
\hline
\raisebox{-1.5em}[0em][0em]{$M_H/\mathrm{GeV}$} &
\raisebox{-1.5em}[0em][0em]{$\sigma_0|_{G_\mu}/\mathrm{pb} $} & 
\raisebox{-1.5em}[0em][0em]{$\sigma^{\rm QCD}_{\rm NLO}/\mathrm{pb} $} & 
\raisebox{-1.5em}[0em][0em]{$\sigma^{\rm QCD+EW}_{\rm NLO}/\mathrm{pb} $} & 
\multicolumn{2}{|c|}{$\sigma^{\rm QCD+EW}_{\rm NLO}/\mathrm{pb}\rule[-3mm]{0mm}{9mm}  $}
\\
\cline{5-6}
 &
 & 
 & 
 & 
$\mu_R = 5\mu_0$ & $\mu_R = \mu_0/5$\\[-2mm]
 &
 & 
 & 
 & 
$\mu_F = \mu_0/5$ & $\mu_F = 5\mu_0$\\[0.5mm]
\hline \hline
80.00	& 2.457(1)	& 2.974(1)	& 2.857(1)	& 2.578(2)	& 3.018(3)  \\ \hline
100.00	& 1.317(1)	& 1.605(1)	& 1.539(1)	& 1.407(1)	& 1.629(1)  \\ \hline
120.00	& 0.7630(1)	& 0.9346(3)	& 0.8947(3)	& 0.8271(2)	& 0.9462(6) \\ \hline
140.00	& 0.4684(2)	& 0.5768(2)	& 0.5508(2)	& 0.5138(1)	& 0.5830(3) \\ \hline
170.00	& 0.2456(1)	& 0.3045(1)	& 0.2900(1)	& 0.2736(1)	& 0.3068(2) \\ \hline
190.00	& 0.1670(1)	& 0.2078(1)	& 0.1978(1)	& 0.1879(1)	& 0.2094(1) \\ \hline
\end{tabular}}
\caption{\label{tab:totzhlhc}Total cross section for 
 $pp\to Z H + X$ at the LHC ($\sqrt{s}=14\;\mathrm{TeV}$)
 in LO, NLO QCD, and including NLO QCD and electroweak corrections in
 the $\GF$-scheme. }

\end{table}

\vspace{1cm}


\begin{table}[p]

\centerline{
\begin{tabular}{|c||c|c|}
\hline 
$M_H/\mathrm{GeV}$ &
CTEQ6M~\cite{Pumplin:2002vw} & 
MRST2001~\cite{Martin:2002aw} \\ \hline \hline
80.00	& 0.5404(2)  $\pm$ 0.021	& 0.5448(2)  $\pm$  0.0097  \\ \hline
100.00	& 0.2685(1)  $\pm$ 0.011	& 0.2698(1)  $\pm$  0.0052  \\ \hline
120.00	& 0.1436(1)  $\pm$ 0.0060	& 0.1437(1)  $\pm$  0.0030  \\ \hline
140.00	& 0.08092(1) $\pm$ 0.0035	& 0.08065(1) $\pm$  0.0018 \\ \hline
170.00	& 0.03679(1) $\pm$ 0.0017	& 0.03644(1) $\pm$  0.00091 \\ \hline
190.00	& 0.02294(1) $\pm$ 0.0011	& 0.02262(1) $\pm$  0.00060 \\ \hline
\end{tabular}}
\caption{\label{tab:pdfwhtev} Parton distribution function (PDF) 
 uncertainties: Total cross section for $p\bar{p} \to W^{\pm} H + X$ at
 the Tevatron ($\sqrt{s}=1.96\;\mathrm{TeV}$) including NLO QCD and
 electroweak corrections in the $\GF$-scheme for different sets of
 parton distribution functions. The results include an estimate of the
 uncertainty due to the parametrization of the parton densities as
 obtained with the CTEQ6~\cite{Pumplin:2002vw} and
 MRST2001~\cite{Martin:2002aw} eigenvector sets (columns 2 and 3,
 respectively). The renormalization and the factorization scales have
 been set to the invariant mass of the Higgs--vector-boson pair, $\mu
 = \mu_0 = \sqrt{s_{VH}}$.}

\vspace{1cm}

\centerline{
\begin{tabular}{|c||c|c|}
\hline 
$M_H/\mathrm{GeV}$ &
CTEQ6M~\cite{Pumplin:2002vw} & 
MRST2001~\cite{Martin:2002aw} \\ \hline \hline
80.00	& 0.3070(1)  $\pm$ 0.012	& 0.3090(1)  $\pm$ 0.0039  \\ \hline
100.00	& 0.1589(1)  $\pm$ 0.0064	& 0.1596(1)  $\pm$ 0.0020  \\ \hline
120.00	& 0.08820(1) $\pm$ 0.0036	& 0.08840(1) $\pm$ 0.0011  \\ \hline
140.00	& 0.05148(1) $\pm$ 0.0021	& 0.05151(1) $\pm$ 0.00066 \\ \hline
170.00	& 0.02472(1) $\pm$ 0.0010	& 0.02469(1) $\pm$ 0.00033 \\ \hline
190.00	& 0.01573(1) $\pm$ 0.00068	& 0.01568(1) $\pm$ 0.00021 \\ \hline
\end{tabular}}
\caption{\label{tab:pdfzhtev}PDF uncertainties: Total cross section for 
 $p\bar{p} \to Z H + X$ at the Tevatron ($\sqrt{s}=1.96\;\mathrm{TeV}$)
 including NLO QCD and electroweak corrections in
 the $\GF$-scheme for different sets of parton distribution
functions.}

\end{table}



\begin{table}[p]

\centerline{
\begin{tabular}{|c||c|c|}
\hline 
$M_H/\mathrm{GeV}$ &
CTEQ6M~\cite{Pumplin:2002vw} & 
MRST2001~\cite{Martin:2002aw} \\ \hline \hline
80.00	& 5.423(2)  $\pm$ 0.18 	& 5.509(2)   $\pm$ 0.071  \\ \hline 
100.00	& 2.859(1)  $\pm$ 0.096	& 2.910(1)   $\pm$ 0.035  \\ \hline 
120.00	& 1.633(1) $\pm$ 0.055	& 1.664(1)  $\pm$ 0.021  \\ \hline 
140.00	& 0.9892(3) $\pm$ 0.034	& 1.010(1)  $\pm$ 0.012 \\ \hline 
170.00	& 0.5078(1) $\pm$ 0.018	& 0.5193(1)  $\pm$ 0.0063 \\ \hline 
190.00	& 0.3466(1) $\pm$ 0.012	& 0.3547(2) $\pm$ 0.0043 \\ \hline 
\end{tabular}}
\caption{\label{tab:pdfwhlhc}PDF uncertainties: Total cross section for 
 $p p \to W^{\pm} H + X$ at the LHC ($\sqrt{s}=14\;\mathrm{TeV}$)
 including NLO QCD and electroweak corrections in
 the $\GF$-scheme for different sets of parton distribution
functions.} 

\vspace{1cm}

\centerline{
\begin{tabular}{|c||c|c|}
\hline 
$M_H/\mathrm{GeV}$ &
CTEQ6M~\cite{Pumplin:2002vw} & 
MRST2001~\cite{Martin:2002aw} \\ \hline \hline
80.00	& 2.857(1)  $\pm$ 0.095  & 2.936(1)   $\pm$ 0.036  \\ \hline
100.00	& 1.539(1)  $\pm$ 0.051	 & 1.583(1)  $\pm$ 0.019  \\ \hline
120.00	& 0.8947(3) $\pm$ 0.030  & 0.9217(3)  $\pm$ 0.011  \\ \hline
140.00	& 0.5508(2) $\pm$ 0.019  & 0.5681(2)  $\pm$ 0.0067 \\ \hline
170.00	& 0.2900(1) $\pm$ 0.010	 & 0.2994(1) $\pm$ 0.0036 \\ \hline
190.00	& 0.1978(1) $\pm$ 0.0069 & 0.2045(1) $\pm$ 0.0025 \\ \hline
\end{tabular}}
\caption{\label{tab:pdfzhlhc}PDF uncertainties: Total cross section for 
 $p p \to Z H + X$ at the LHC ($\sqrt{s}=14\;\mathrm{TeV}$)
 including NLO QCD and electroweak corrections in
 the $\GF$-scheme for different sets of parton distribution
functions.}

\end{table}


\end{document}